\documentclass[aps,preprint,epsfig,rotate]{revtex4}
\usepackage{sansmath,pdflscape,graphicx,rotating}
\usepackage{tikz}
\usetikzlibrary{calc,arrows,decorations.pathmorphing,intersections}
\usepackage[font={small,sf},labelfont={bf},labelsep=endash]{caption}
\begin{document}
%\begin{doublespace}
\title{Bound state spectra and properties of the doublet states in three-electron 
       atomic systems} 

 \author{Alexei M. Frolov}
 \email[E--mail address: ]{afrolov@uwo.ca}

\affiliation{Department of Applied Mathematics \\
 University of Western Ontario, London, Ontario N6H 5B7, Canada}

\author{Mar\'{\i}a Bel\'en Ruiz}
\email[E--mail address: ]{maria.belen.ruiz@fau.de}

\affiliation{Department of Theoretical Chemistry, \\
Friedrich-Alexander-University Erlangen-N\"urnberg, Egerlandstra\ss e 3, 
D-91058, Erlangen, Germany}

\author{David M. Wardlaw}
 \email[E--mail address: ]{dwardlaw@mun.ca}

\affiliation{Department of Chemistry, Memorial University of Newfoundland, St. John's, 
             Newfoundland and Labrador A1C 5S7, Canada}

\date{\today}

\begin{abstract}

The bound state spectra of the doublet states in three-electron atomic systems are investigated. By using different 
variational expansions we determine various bound state properties in these systems. Such properties include the 
electron-nucleus and electron-electron delta-functions and cusp values. The general structure of the bound state 
spectra in several three-electron atomic systems (Li, Be$^{+}$, C$^{3+}$ and F$^{6+}$) is investigated with the use 
of the Hylleraas-Configuration Interaction and the Configuration Interaction wave functions. The advantage of our 
Configuration Interaction based procedure is that it provides high numerical accuracy for all 
rotationally excited states, including the bound states with $L \ge 7$. 

%PACS number(s): 03.65.Ge, 31.15.ac and 31.15.xf
\end{abstract}

\maketitle
%\noindent \vspace{0.2in}

\section{Introduction}\label{intro}

In this study we investigate the bound doublet states in the three-electron atoms and ions by using a few different 
variational expansions written in the relative coordinates, or interparticle distances $r_{ij} =  \mid {\bf r}_i - 
{\bf r}_j \mid = r_{ji}$, where ${\bf r}_i$ ($i$ = 1, 2, 3) are the Cartesian coordinates of the three electrons, while
${\bf r}_4$ are the Cartesian coordinates of the positively charged nucleus. Another our goal is to determine the bound 
state properties of some of these bound (doublet) states and discuss problems arising during this procedure. By calculating 
these properties we want to correct mistakes which have been made in earlier papers and continue to propagate in the modern 
scientific literature.

As is well known in the lowest-order approximation upon the fine structure constant $\alpha (= \frac{e^2}{\hbar c}$) an
arbitrary three-electron atom/ion is described by the non-relativistic Schr\"{o}dinger equation $H \Psi = E \Psi$, where $E < 
0$ and the bound state wave function $\Psi$ has the unit norm. The non-relativistic Hamiltonian $H$ of the three-electron 
atom/ion is (see, e.g., \cite{LLQ})
\begin{eqnarray}
 H = -\frac{\hbar^2}{2 m_e} \Bigl[\nabla^2_1 + \nabla^2_2 + \nabla^2_3 + \frac{m_e}{M} \nabla^2_4 
 \Bigr] - \frac{Q e^2}{r_{14}} - \frac{Q e^2}{r_{24}} - \frac{Q e^2}{r_{34}} + \frac{e^2}{r_{12}} 
 + \frac{e^2}{r_{13}} + \frac{e^2}{r_{23}} \label{Hamil}
\end{eqnarray}
where $\hbar = \frac{h}{2 \pi}$ is the reduced Planck constant, $m_e$ is the electron mass and $e$ is the electric charge 
of an electron. In this equation and everywhere below in this study the subscripts 1, 2, 3 designate the three electrons 
$e^-$, while the subscript 4 denotes the heavy nucleus with mass $M$ ($M \gg m_e$) and positive electric (nuclear) charge 
$Q e$. The notations $r_{ij} = \mid {\bf r}_i - {\bf r}_j \mid = r_{ji}$ stand for the six interparticle distances (= 
relative coordinates) defined above. In Eq.(\ref{Hamil}) and everywhere below in this work we shall assume that $(ij)$ 
= $(ji)$ = (12), (13), (14), (23), (24), (34). Below only atomic units $\hbar = 1, \mid e \mid = 1, m_e = 1$ are employed. 
In these units the explicit form of the Hamiltonian $H$, Eq.(\ref{Hamil}), is simplified
\begin{eqnarray}
 H = -\frac12 \Bigl[ {\bf p}^2_1 + {\bf p}^2_2 + {\bf p}^2_3 + \frac{1}{M} {\bf p}^2_N \Bigr] - \frac{Q}{r_{14}} - 
 \frac{Q}{r_{24}} - \frac{Q}{r_{34}} + \frac{1}{r_{12}} + \frac{1}{r_{13}} + \frac{1}{r_{23}} \label{Hamil1}
\end{eqnarray}
where the notations ${\bf p}_i$ designate the momenta of the electrons ($i$ = 1, 2, 3) and the nucleus ($i$ = 4, or $i = N$). In 
atomic units we have ${\bf p}^{2}_{i} = p^{2}_{i} = -\nabla^{2}_{i}$. 

For last fifty years (since the paper \cite{Lars} was published) the problem of highly accurate calculations of the bound states in 
three-electron atomic systems has attracted a significant attention. A very nice review of the three-electron atomic problem can be
found in \cite{McW} which also contains all references prior to 1969. Recently, impressive calculations have been achieved for the low-lying states 
of three-electron systems using Hylleraas-type wave functions \cite{Ho,Clary,Sims,YD}. Such calculations are the result 
of developments in the methods of evaluation of the three-electron integrals. The integral methods used to perform 
the calculations of this work are described in Refs. \cite{Sims3e,Ruiz3e,Fro2003}. The rotationally excited states or 
Rydberg states of three-electron systems have attracted considerable 
attention and calculations have been done using theoretical methods to describe the core two-electron ion and the weakly 
bound excited electron \cite{Bhatia,Chen}. 
A great theoretical interest in three-electron atomic systems can 
be explained by a number of facts known for such systems. First, three-electron
atomic systems are very convenient objects for the study of the overall and relative contributions of the electron-electron correlations. It
follows from the fact that the potential energy in the Hamiltonian, Eq.(\ref{Hamil}), is represented by two sums each of which 
contains equal number of terms: electron-nucleus attractions and electron-electron repulsions. Therefore, by varying only one parameter 
$Q$ in Eq.(\ref{Hamil}) we can study electron-electron correlations in such systems and relations between electron-electron and 
electron-nucleus correlations. Second, all three-electron atoms and ions are well known subjects in atomic physics. On the 
other hand, the overall accuracy of bound state computations achieved for three-electron atoms and ions is still substantially 
lower than analogous accuracy known for two-electron atomic systems. Therefore, the development of new, rapidly converging 
variational expansions and effective, fast computational methods is of paramount importance for the future of highly accurate 
calculations of three-electron atoms and ions. Third, despite a rapidly increasing stream of computational publications on three-electron 
atomic systems some fundamental aspects of such systems have not been discussed yet. For instance, the quality of the wave 
function constructed for an arbitrary Coulomb few- and many-body system can be tested by comparing the computed and expected cusp 
values, or cusps, for short. However, it appears that neither the electron-nucleus, nor electron-electron cusps were ever 
determined for three-electron atoms and ions. Another well known problem is related to an incorrect definition of some electron-nucleus 
and electron-electron properties accepted in quite a few earlier works performed for three-electron atoms and ions. Furthermore, it
was found that the expectation values of some bound state properties computed with the use of different variational expansions deviate 
from each other. In a few cases such deviations are relatively large and may lead to contradictions with the known experimental 
results. 
 
In our analysis below we re-consider all these problems by performing highly accurate calculations of some bound states in 
three-electron atomic systems. This work has the following structure. In the next Section we discuss the semi-exponential 
variational expansion in relative coordinates \cite{Fro2010}. This very compact and accurate variational expansion was introduced 
in \cite{Fro2010}. In this study this expansion is applied to determine various bound state properties in 
the ground $2^2S-$state(s) of the three-electron Li-atom and Be$^{+}$ and B$^{2+}$ ions. The computed expectation values include 
electron-nucleus and electron-electron delta-functions and cusp-values. In Section III we consider a number of bound (doublet) 
${}^2S-$states in different three-electron atoms and ions by using another variational expansion of the wave functions. Here all 
wave functions are approximated by expansions in six-dimensional gaussoids \cite{KT}. Note that analytical 
formulas for all matrix elements which arise in variational calculations of arbitrary $A-$body system in the basis of 
multi-dimensional gaussoids, where $A \ge 2$ is an arbitrary integer, have been derived in the mid-1970's (see \cite{KT} and 
earlier references therein). Subsequently this variational expansion was successfully applied for bound state computations in many 
hundreds of nuclear, atomic and molecular systems. In Section III we use this variational expansion to determine the total energies 
and other bound state properties of some excited states in the Li-atom. The results obtained in  Section III can also be used to 
correct a few mistakes and misprints made in earlier studies of three-electron atomic systems. In Section IV by using the 
Hylleraas-Configuration Interaction (Hy-CI) and Configuration Interaction (CI) 
variational expansions of the wave functions we investigate the general structure of the bound state spectra (or optical spectra, for 
short) of three-electron (or Li-like) atoms and ions. Section IV also contains a brief discussion of the `scaling' procedure which can 
be used to study the electron-electron correlations in three-electron atomic systems. Concluding remarks can be found in the last Section.
 
\section{Semi-exponential variational expansion}

A semi-exponential variational expansion for three-electron wave functions was developed a few years ago \cite{Fro2010} in order to modify 
the traditional Hylleraas variational expansion which has been used for three-electron atomic systems since Larsson's work \cite{Lars}. 
Recently it became clear that the Hylleraas variational expansion of the three-electron wave functions has a very slow convergence rate. 
Our main idea in \cite{Fro2010} was to increase the total number of varied, non-linear parameters in each basis function included from the 
variational expansion. Finally, we developed a very effective variational expansion for three-electron atoms and ions which can now be used 
to construct very compact and accurate wave functions for arbitrary three-electron atom and/or ion. For simplicity, below we consider only 
the doublet $2^{2}S(L = 0)$-states in the three-electron Li atom and similar Be$^{+}$ and B$^{2+}$ ions. By using our semi-exponential 
variational expansion we determine bound state properties of these three-electron atomic systems. It should be mentioned that some of these 
properties have never been determined in earlier studies. 

The variational wave function of the doublet $S(L = 0)$-states of the three-electron atom/ion is written in the form
\begin{eqnarray}
 \Psi_{L=0} = \psi_{L=0}(A; \bigl\{ r_{ij} \bigr\}) (\alpha \beta \alpha - \beta \alpha \alpha) + \phi_{L=0}(B; \bigl\{ 
 r_{ij} \bigr\}) (2 \alpha \alpha \beta  - \beta \alpha \alpha - \alpha \beta \alpha) \label{psi}
\end{eqnarray}
where $\psi_{L=0}(A; \bigl\{ r_{ij} \bigr\})$ and $\phi_{L=0}(B; \bigl\{ r_{ij} \bigr\})$ are the two independent spatial 
parts (also called the radial parts, or radial functions) of the total wave function. Each of these two radial functions is, in 
fact, a radial factor (for states with $L = 0$) in front of the corresponding three-electron spin functions $\chi_1 = \alpha \beta 
\alpha - \beta \alpha \alpha$ and $\chi_2 = 2 \alpha \alpha \beta  - \beta \alpha \alpha - \alpha \beta \alpha$. Here the notations 
$\alpha$ and $\beta$ are the one-electron spin-up and spin-down functions, respectively (their definition can be found, e.g., in 
\cite{Dir}). The notations $A$ and $B$ in Eq.(\ref{psi}) mean that the two sets of non-linear parameters associated with radial functions $\psi$ 
and $\phi$ can be optimized independently. In the general case, each of the radial basis functions explicitly depends upon all six interparticle 
(relative) coordinates $r_{12}, r_{13}, r_{23}, r_{14}, r_{24}, r_{34}$. It is clear that in actual bound state calculations 
only one spin function, e.g., the $\chi_1$ function, can be used. Note also that when the corresponding radial function has been 
constructed, then one can use an artifice called the `doubling' of the wave function (see, e.g., \cite{Fro2011}). This trick is based on 
the use of the same set of non-linear parameters in the two radial functions in Eq.(\ref{psi}). Obviously, this is not an optimal choice
of the non-linear parameters in the wave function, but in many cases this simple handling leads to a substantial improvement of the answer. 

The semi-exponential variational expansion of the radial function $\psi_{L=0}(A; \bigl\{ r_{ij} \bigr\})$ from Eq.(\ref{psi}) is written in 
the form
\begin{eqnarray}
 \psi_{L=0}(A; \bigl\{ r_{ij} \bigr\}) = \sum^N_{k=1} C_k r^{n_1(k)}_{23} r^{n_2(k)}_{13} r^{n_3(k)}_{12} r^{m_1(k)}_{14} 
 r^{m_2(k)}_{24} r^{m_3(k)}_{34} exp(-\alpha_{k} r_{14} -\beta_{k} r_{24} -\gamma_{k} r_{34}) \label{semexp}
\end{eqnarray}
where $\alpha_k, \beta_k, \gamma_k$ ($k = 1, 2, \ldots, N$) are the varied non-linear parameters. The presence of the varied non-linear 
parameters in Eq.(\ref{semexp}) is the main and very important difference with the traditional Hylleraas variational expansion (see, e.g., 
\cite{Lars}). In the last case in Eq.(\ref{semexp}) we have $\alpha_1 = \ldots = \alpha_N, \beta_1 = \ldots = \beta_N$ and $\gamma_1 = 
\ldots = \gamma_N$. Note that all matrix elements of the Hamiltonian, Eq.(\ref{Hamil}), and overlap matrix needed in computations with 
the use of the semi-exponential basis, Eq.(\ref{semexp}), contain the same three-electron integrals which arise for the usual Hylleraas 
expansion (for more detail, see discussion in \cite{Fro2010}). In other words, numerical calculation of all matrix elements with 
semi-exponential functions is not a more difficult task than for the traditional Hylleraas radial functions. This also simplifies numerical 
computations of the bound state properties (i.e. expectation values) in the semi-exponential basis set. In particular, our algorithms used in 
calculations of all required matrix elements is based on the old Perkins formula for three-electron integrals \cite{Per} in relative 
coordinates. The explicit symmetrizaton of the wave function upon all permutations of identical particles (electron) is discussed in detail 
in our earlier works (see, e.g., \cite{Fro2010} and references therein). Note also that all calculations in this work have been performed with 
the use of standard quadruple precision accuracy (30 decimal digits per computer word). In our calculations we have used variational wave 
functions for the Li atom, Be$^{+}$ and B$^{3+}$ ions with 60 terms. To construct these wave functions we follow the procedure described in detail
in \cite{Fro2011}. 

As mentioned above in this study we consider the doublet $2^2S(L = 0)$-states in the three-electron Li atom and in analogous Be$^{+}$ and B$^{2+}$ 
ions. The corresponding expectation values (or properties) can be found in Table I (in atomic units $\hbar = 1, e = 1, m_e = 1$). For most of the 
properties mentioned in Table I their meaning is clear from the notation used. Therefore, here we can restrict ourselves to a few following 
remarks. All electron-nucleus properties are designated below with the use of a general index $eN$ (electron-nucleus), while the notation $ee$ 
stands for electron-electron properties. For instance, the notations $\langle r_{eN} \rangle$ and $\langle r_{ee} \rangle$ mean the electron-nucleus 
and electron-electron distances, respectively. The total energies obtained with these 60-term wave functions are: $E$ = -7.47805 86794 4751 $a.u.$ (Li 
atom), $E$ = -14.32475 2251 4351 $a.u.$ (Be$^{+}$ ion) and $E$ = -23.42457 46190 439 $a.u.$ (B$^{2+}$ ion), respectively.

The expectation values of the electron-nucleus and electron-electron delta-functions, i.e. $\langle \delta_{eN} \rangle$ and $\langle \delta_{ee} 
\rangle$ are of great interest in various applications. Therefore, it is important to be sure that these values have been determined correctly. For 
atomic and molecular systems with the Coulomb interaction between each pair of particles there is a very effective test which can be used to estimate 
the actual accuracy of the expectation values of all two-particle delta-functions $\langle \delta({\bf r}_{ij}) \rangle$. This test is based on 
accurate numerical evaluation of the electron-nucleus and elecron-electron cusp values. In general, the cusp value between two point particles $a$ and 
$b$ with electrical charges $q_a$ and $q_b$ and masses $m_a$ and $m_b$ (in atomic units) is:
\begin{eqnarray}
  \overline{\nu}_{ab} = q_a q_b \frac{m_a m_b}{m_a + m_b} \label{cuspe}
\end{eqnarray}
This is the expected (or `classical') numerical value of the cusp between the electrically charged particles $a$ and $b$. In general, the cusp is 
defined by the equation (see, e.g., \cite{cusp})
\begin{eqnarray}
  \nu_{ab} = \frac{\langle \delta({\bf r}_{ab}) \frac{\partial}{\partial r_{ab}} \rangle}{\langle \delta({\bf r}_{ab}) \rangle} \label{cusp}.
\end{eqnarray}
Numerical coincidence of the $\nu_{ab}$ and $\overline{\nu}_{ab}$ values is a good test for the overall quality of the wave functions constructed for
different atomic and molecular systems. 

It should be mentioned that the `classical' definition of the cusp (or cusp-condition) given by Eq.(\ref{cusp}) can be generalized to quantum mechanics 
in a number of different ways. To explain this let us introduce the following cusp-operator 
\begin{eqnarray}
  \hat{\nu}_{ab} = \frac{1}{\langle \delta({\bf r}_{ab}) \rangle} \delta({\bf r}_{ab}) \frac{\partial}{\partial r_{ab}} \label{cusp2}
\end{eqnarray}
For the expectation values one finds $\nu_{ab} = \langle \Psi \mid \hat{\nu}_{ab} \mid \Psi \rangle$. Formally, we can consider the quantity $\langle \Psi_i 
\mid \hat{\nu}_{ab} \mid \Psi_j \rangle$, where $\Psi_i$ and $\Psi_j$ are the two different wave functions from the bound state spectrum. The last 
expression, however, is not symmetric upon the $i \leftrightarrow j$ substitution. Such a symmetry can be restored with the following redefinition of the 
cusp operator, Eq.(\ref{cusp2}):
\begin{eqnarray}
 \hat{\nu}_{ab} = \frac{1}{2 \langle \delta({\bf r}_{ab}) \rangle} \Bigl[ {}_{\leftarrow}\Bigl( \frac{\partial}{\partial r_{ab}}\Bigr) \delta({\bf r}_{ab}) 
 + \delta({\bf r}_{ab}) \Bigl(\frac{\partial}{\partial r_{ab}}\Bigr)_{\rightarrow} \Bigr] \label{cusp3}
\end{eqnarray}
where the differential operator with the index `$\rightarrow$' acts on its right, while analogous operator with the index `$\leftarrow$' acts on its left.
The definition of the cusp operator based on Eq.(\ref{cusp3}) has a number of advantages in applications. In particular, it allows one to obtain more
accurate values of the electron-nucleus and electron-electron cusps with the use of relatively short wave functions. In this study to determine the numerical 
values of all cusps mentioned in this Section we used the cusp operators written in the form of Eq.(\ref{cusp3}).

\section{Basis of six-dimensional gaussoids}

For three-electron atoms and ions there is another variational expansion which can be effective to perform accurate computations of low-lying bound 
states, e.g., the low-lying bound $S(L = 0)$- and $P(L = 1)$-states. The high efficiency of such an expansion for bound state computations is based on 
very simple and numerically stable formulas for all matrix elements needed in computations of the total energies of these states. This variational 
expansion was developed in the mid-1970 by Kolesnikov and his group (see, e.g., \cite{KT} and references therein). In general, this expansion 
is also represented in the form of Eq.(\ref{psi}), but only radial functions are represented as linear combinations of the six-dimensional gaussoids, e.g., 
for $\psi_{L=0}(A; \bigl\{ r_{ij} \bigr\})$ we have
\begin{eqnarray}
 \psi_{L=0}(A; \bigl\{ r_{ij} \bigr\}) = \sum^N_{k=1} C_k exp(-\alpha^{(k)}_{12} r^2_{12} -\alpha^{(k)}_{13} r^2_{13} -\alpha^{(k)}_{23} r^2_{23}
 -\alpha^{(k)}_{14} r^2_{14} -\alpha^{(k)}_{24} r^2_{24} -\alpha^{(k)}_{34} r^2_{34}) \label{gauss}
\end{eqnarray}
where $C_k$ are the linear coefficients (or linear variational parameters), and $\alpha^{(k)}_{ij}$ are the varied non-linear parameters. It is clear 
that for Coulomb systems each of the basis functions in Eq.(\ref{gauss}) does not provide the correct long-range asymptotic between particles. Indeed, for the 
$i$ and $j$ particles the leading term in the asymptotics of the actual atomic wave function must be represented in the form $r^{K}_{ij} \exp(-\beta r_{ij})$. 
Additional terms for Coulomb (or atomic) systems may include different powers of $log r_{ij}, log \mid log r_{ij} \mid$ and other similar factors. This follows 
from the general theory of bound states in the few-body Coulomb systems (see, e.g., \cite{Fock1954} and references therein). It is clear that such terms cannot 
be derived from the approximate wave function Eq.(\ref{gauss}). Based on this conclusion we can predict a number of different bound state properties with the 
use of variational expansion, Eq.(\ref{gauss}). In reality, the variational expansion, Eq.(\ref{gauss}), can be used for approximate evaluations of the bound 
state properties in atomic systems.  

Formally, this means that the expectation values determined with the use of multi-dimensional gaussoids can be used as a one-sided approximation to the actual 
bound state properties of atomic systems. The procedure works in the following way. By using relatively large numbers of multi-dimensional gaussoids with 
carefully optimized non-linear parameters $\alpha^{(k)}_{ij}$ in Eq.(\ref{gauss}) we can obtain accurate numerical values of the total energies. Then we continue
optimization of the non-linear parameters $\alpha^{(k)}_{ij}$ in Eq.(\ref{gauss}). The total energies continue to converge to their limiting values. However, 
the expectation values of many bound state properties rapidly improve during such an additional optimization. The principal question here is to evaluate the 
convergence rate of the variational expansion, Eq.(\ref{gauss}). In our calculations we have found that the actual convergence rate for the total energies is 
relatively fast and already for $N = 700 - 1000$ in Eq.(\ref{gauss}) we obtain results which can be considered as accurate, or even highly accurate. 

The total energies of the ground $2^2S(L = 0)$ and some excited $n^2S(L = 0)$-states ($n \ge 3$) in some three-electron atoms and ions can be found in Table II
(in atomic units). Table III contains the bound state properties determined with the use of our best variational wave functions. In such calculations we have 
used only 700 basis functions (six-dimensional gaussoids) with carefully optimized non-linear parameters. It should be mentioned here that highly accurate 
wave functions can be constructed with the use of 2000 (or more) basis functions (six-dimensional gaussoids) with carefully optimized non-linear parameters. 
The explicit construction of such accurate wave functions with 2000 terms (or more) does not require excessive computational resources, but in calculations 
performed for this study we restrict ourselves to bound state wave functions with $N = 700$ terms only. There are two reasons for this. First, we compare our 
bound state properties with analogous properties computed in earlier papers (see, e.g., \cite{King91} and references therein) where the total number of basis 
functions used was $\approx 500 - 800$, which is comparable to 700. Second, many differences between accurate bound state properties obtained with $N = 700$ 
basis functions in Eq.(\ref{gauss}) and highly accurate values determined with $N = 2000$ basis functions in Eq.(\ref{gauss}) are of interest for some special 
purposes only. The overall accuracy of our computed bound state properties is sufficient to predict the results of various experiments performed for three-electron 
atomic systems. Moreover, our results obtained for the three-electron atomic systems presented in Tables I and III can be used as a basis for more accurate 
calculations of bound state properties in the future. Note that the expectation values of the two-particle cusps cannot be evaluated with the use of the 
KT-expansion, Eq.(\ref{gauss}). Formally, for these values one always finds zero, if the variational KT-expansion is used in the cusp definitions, Eqs.(\ref{cusp}) - 
(\ref{cusp3}). Theoretically, these expectation values can be corrected, e.g., by using different representations of the wave functions at the short interparticle
distances. The linear coefficients in such a `new' representation of the wave function are determined from the known KT-expansions in some `intermediate' areas. 
In reality, very often this procedure does not lead to the accurate cusp values. Furthermore, the answer substantially depends upon the size of this `intermediate' 
area. Therefore, the comparison of the `expected' and actual cusp values can be made only approximately, i.e. the original sense of the cusp condition as a simple 
(but effective) criterion for the wave functions is essentially lost.        

To conclude this Section we note that there is another fundamental question about accurate calculations of bound state properties. In general, the physical 
meaning of each property is clear only for those few-body systems where any group of particles do not form any closed shells, or cluster structures. Otherwise, the 
meaning of the computed properties rapidly became unclear when the total number of identical particles increases. For instance, for highly excited ${}^{2}S(L = 
0)$-states in the Li atom (see Tables II and III) one of the three electrons is moving as almost free particle. In old atomic physics such a motion was considered as 
the pre-dissociation. This means that the electron-nucleus distance for that electron is extremely large in comparison to the analogous distances determined for 
two other electrons which occupy the closed $1s^2$-shell. To calculate the expectation value $\langle r_{eN} \rangle$ we need to sum up all three electron-nucleus 
distances (two relatively short and one very large $r_{eN}$ distances) and divide the arising sum by three. Finally, we obtain the expectation value $\langle 
r_{eN} \rangle$ which, in fact, gives us no useful information. It is similar to the mean patient temperature measured over the `whole hospital'. The same is true for other 
bound state properties, including all electron-nucleus and electron-electron properties. Formally, in such cases we need to assume that some originally identical 
particles becomes `less-identical', or even `non-identical' for highly excited states. If such an assumption has been made, then the expectation values of different 
properties provide us with useful information which has a direct physical sense. However, at this moment we do not have reliable recipes for calculations of such 
values, since $a$ $priori$ the `power of non-identity' for the outer most electron is no clear, e.g., in the $4^2S-, 5^2S-$ and $6^2S-$state (see Tables II and 
III). The power of `non-identity' increases with the excitation, i.e. with the principal quantum number $n$.       

\section{On general structure of the bound state spectra}

The general structure of the bound state spectra in three-electron atoms and ions was considered in our earlier paper \cite{Bel1}. In this paper by using the energy 
values obtained in \cite{Bel1} for the Li atom and the Be$^+$ ion we have drawn the spectral diagrams of these species. For this study we slightly improved numerical
values of the total energies and draw the spectral diagrams for the doublet series of the Li atom and the Be$^+$ ion (see Figs. 1 and 2). In particular, the spectra of 
the Li atom is identical to the experimental spectra given in the classical book \cite{Kik}. In this work we would like to continue our theoretical analysis of the bound 
state spectra in three-electron atoms and ions. The analysis of the bound state spectra in similar atomic systems is based on the use of fast-convergent variational 
expansions which provide high numerical accuracy not only for the ground and some low-lying rotationally excited states, but for all excited states in few-electron atom(s) 
and ions. This expansion is the Hylleraas-Configuration Interaction (Hy-CI) wave function, proposed by Sims and Hagstrom \cite{SH1,SH2}, which we have used to determine the 
$S$-, $P$-, $D$-states of the C$^{3+}$ and F$^{6+}$ ions and the Configuration Interaction (CI) wave function with Slater orbitals and $LS$ configurations for the states of 
higher symmetry $F$, $G$, $H$, $I$, $K$, $L$, $M$, $N$ and states with $L=20$. Note that the Hy-CI is general for any symmetry. At the moment in our computer program the kinetic energy integrals are restricted 
to $l = 3$, work is in progress to generalize our code for $l \ge 3$. Nevertheless, the CI wave function leads to very good results for states with quantum number $L \ge 3$. 
The Hy-CI and CI wave functions can be summarized in the following expression:
\begin{equation}
\Psi =\sum_{p=1}^NC_p\Phi _p,\qquad \Phi _p=\hat{O}(\hat{L}^2)\hat{\mathcal{A}}\psi _p\chi 
\end{equation}
$N$ is the number of configurations $\Phi_p$, and $C_p$ a variational coefficient. All configurations are symmetry-adapted (this is expressed in the last equation with 
the operator $\hat{O}(\hat{L}^2)$). The operator $\hat{\mathcal{A}}$ is the antisymmetrization operator and $\chi$ is the spin eigenfunction: 
\begin{equation}
\chi =\left[ (\alpha \beta -\beta \alpha )\alpha \right]. 
\end{equation}
The Hartree products are multiplied by up to one interelectronic coordinate $r_{ij}$ 
\begin{equation}
\psi _p=r_{ij}^\nu \prod_{k=1}^n\phi _k(r_k,\theta _k,\varphi _k),
\end{equation}
where the choice $\nu = 0$ corresponds to the CI wave function, while for $\nu = 1$ we have the Hy-CI wave function. To calculate the bound (doublet) states in the 
C$^{3+}$ and F$^{6+}$ ions we have optimized the exponents of the wave function expansions for the Li and Be$^{+}$ atoms of Ref.\cite{Bel1} for the nuclear charges 
$Z = 6$ and $Z = 9$. The Hy-CI calculations have been conducted using quadruple precision arithmetic and we have employed $\approx$ 1000 configurations, while in the CI calculations 
double precision has been used and we have employed between 1000 and 1500 configurations.
The construction of the configurations is described in \cite{Bel1}. The configurations of the newly calculated states in this work are similar to the ones of \cite{Bel1} employing higher 
$l$ quantum number for the outer electron. The CI procedure is very stable for calculations with high quantum number L. This is based in the very stable two-electron integral algorithms used in the calculations. These 
algorithms have been described in detail by Eqs. (A.33-A.35) of Ref. \cite{Ruiz2e}. The accuracy of the CI calculations is $\approx 1$ mhartree ($1\times 10^{-3} a.u.$).     
The obtained energies are listed in Table IV. We can compare the energy values of the $S$-states obtained by the gaussoids expansion in Table II with the ones by the Hy-CI method.
All the energy values are very close, the differences are in the order of few microhartrees ($1\times 10^{-6} a.u.$). The ground state energies of the Li atom and Li-like ions are slightly better by using the 
Gaussoids expansion, while the low-lying excited $P$-states are better by using the Hy-CI method. For the Li atom and Be$^+$ ion we compared the energies of low-lying states of $S$-, $P$- 
and $D$-symmetry in \cite{Bel1} obtaining differences in the order of few microhartrees.    
In this work we have calculated higher states of the Li atom and Be$^{+}$ ion using the CI method. 
For the ions, the literature is very scarce. We were able to compare the ground and lowest $P$-symmetry states of the C$^{3+}$ and F$^{6+}$ ions 
with Hylleraas-type calculations \cite{YTD}, see Table V. Note that the Hy-CI energy values of the $P$-states are very accurate, showing that the Hy-CI wave function is 
very advantageous for the calculation of states with non-zero angular momentum. 
With this method the accuracy of higher excited states of the lowest symmetries (like it is the case of $8P$, $8D$ which are not listed in Table IV) is lower than the accuracy 
of low-lying excited states of higher symmetry, i.e. $8F$, $8G$, \ldots , $8K$. The reason is the state $8K$ is the first solution of the eigenvalue equation for a wave function of K-symmetry, 
while i.e. $8P$ is the seventh solution of its corresponding Schr\"odinger equation. In this respect, one can observe innacuracies in the obtained energy of the weakly bound Rydberg states.   

With the data of Table IV we have drawn the spectral diagrams of these ions, see Figs. 3 and 4.
We have scaled these diagrams taking for every one the ground state energy level as lowest point and the limit of inonization as highest point, and calculating the position
of the states with respect to this interval. Therefore we can compare the relative position or contraction of the energy levels of every specie with respect to its ground state. Our theoretical distribution of energy levels agrees completely with the experimental results \cite{NIST}. 
In addition we have determined in this work atomic levels which experimental values have not been yet reported, like states with $ L \ge 4$ in Li atom, $L \ge 6$ of Be$^+$ 
ion, $L \ge 7$ of C$^{3+}$ and $L \ge 4$ of F$^{6+}$ ion.  
Using theoretical and experimental results we can conclude that the order of the energy levels in the duplet states of the neutral atom and isoelectronic ions is
$S <   P <   D <   F <   G <   H <   I <   K <   L <   M <   N <   \dots $

As is well known from basics of atomic spectroscopy (see, e.g., \cite{Sob}) all three-electron atoms and ions are observed in the two series of states: (a) doublet 
states with $S_e = \frac12$, and (b) quartet states with $S_e = \frac32$, where $S_e$ designates the total electron spin. The quartet states in these atomic 
systems are non-stable, i.e. they decay for relatively short times. Thus, the only bound states which are observed in actual spectroscopy of three-electron 
atomic systems are the doublet states. Therefore, below in this Section we restrict ourselves to the consideration of the doublet (bound) states only. As 
mentioned in the Introduction the atomic Hamiltonian, Eq.(\ref{Hamil}), contains equal numbers of electron-nucleus and electron-electron terms. Indeed, in 
Eq.(\ref{Hamil}) one finds three terms which describe electron-nucleus attraction and three other terms which describe electron-electron repulsion. The first
four terms in  Eq.(\ref{Hamil}) represent the kinetic energy of the four particles (three electrons and one nucleus). This means that by varying the electric 
charge of the central nucleus $Q$ we can study the role of electron-electron correlations in three-electron atomic systems. It is clear that by increasing 
$Q$ we can reduce the overall role of electron-electron repulsions in three-electron ions. There are three electron-nucleus terms and three electron-electron
terms in the potential energy of the Hamiltonian, Eq.(\ref{Hamil}). This means that by varying $Q$ we change relations between each electron-nucleus and 
electron-electron terms. On the other hand, an analogous relation between the total sum of the three electron-nucleus terms and the total sum of the three 
electron-electron terms changes in the same proportion.        

The electronic structure of the ground (bound) doublet $2^2S$-state of the three-electron Li atom is $1s^2 2s^1$, while all excited states have a similar structure 
where the two electrons occupy the $1s^2$-electron shell (its excitation energy is extremely large), while the third electron can occupy any free electron orbital 
in the atom. Formally, we can say that the third electron is located in one of the $n\ell$-shells, where $\ell \ge 0$, $n = k + \ell + 1$ and $k \ge 0$ (all these 
numbers are integer). Possible excitations of the Li atom always mean the excitation of the third electron, which sometimes is considered as an `optical' electron. 
Any excitation of the central electron $1s^2$-shell leads to the complete dissociation of the whole atom. In the lowest-order approximation we can say that the 
optical spectrum of the Li-atom is similar to the well known spectrum of the hydrogen atom. However, the actual similarities in optical spectra can be found only for 
highly excited bound states in the Li atom which are often called the Rydberg states. 

As follows from the general theory of atomic spectra the total energies of weakly-bound Rydberg states in any neutral atom must be represented by a formula which
is similar to the well known formula for the hydrogen-like atoms. Let us discuss such a formula in detail. First, note that the dissociation threshold for the 
neutral Li-atom corresponds to the formation of the two-electron Li$^{+}$ ion in its ground $1^1S$-state (singlet). The non-relativistic energy of this state is
$E_{{}^{\infty}{\rm Li}^{+}}$ $\approx$ -7.279913 412669 305964 91895(15) $a.u.$ \cite{Fro2014}. This dissociation threshold corresponds to the following 
ionization process for the neutral Li atom
\begin{eqnarray}
  {\rm Li} = {\rm Li}^{+}(1^1S) + e^{-} \label{Rydb0}
\end{eqnarray}
where the symbol Li$^{+}(1^1S)$ means that the final two-electron Li$^{+}$ ion is in its ground (singlet) $1^1S$-state. Now, we can write the following expression for 
the total energies of the weakly-bound states, i.e. for the states which are close to the dissociation threshold of the Li atom (in atomic units):
\begin{eqnarray}
  E({\rm Li}; n L) = E({\rm Li}^{+}; 1^1S) - \frac{m_e e^4}{2 \hbar^2} \frac{1 - \epsilon_{\ell}}{(n + \Delta_{\ell})^2} = 
  -7.279913 412669 305\ldots - \frac{1 - \epsilon_{\ell}}{2 (n + \Delta_{\ell})^2} \label{Rydb}
\end{eqnarray}
where $L = \ell$ (in this case), $n$ is the principal quantum number of the $n L$ state ($L$ is the angular quantum number) of the Li atom, while $\epsilon_{\ell}$
and $\Delta_{\ell}$ are the Rydberg corrections ($\Delta_{\ell}$ is also called the `quantum defect') which explicitly depends upon $\ell$ (angular momentum of the 
outer most electron) and the total electron spin of this atomic state. It can be shown that both Rydberg corrections rapidly vanish when $\ell$ increases (for 
given $n$ and $L$). Moreover, these two corrections also decrease when the principal quantum number $n$ grows. The formula, Eq.(\ref{Rydb}), can be used 
to approximate the total energies of weakly-bound Rydberg states in the Li atom. In reality, by using a few accurate (or highly accurate) results from numerical 
calculations of some excited (bound) states in the Li atom one finds the approximate values for the $\epsilon_{\ell}$ and $\Delta_{\ell}$ constants in Eq.(\ref{Rydb}). 
Analogous formulas can be derived to describe the total energies of the excited bound states in three-electron ions, e.g., in the Be$^{+}$, C$^{3+}$ and F$^{6+}$ ions. 
However, after neon the validity of the non-relativistic approximation for three-electron ions rapidly diminishes as the parameter $Q$ in Eq.(\ref{Hamil}) continue to 
grow. In the lowest-order approximation the leading relativistic corrections can directly be introduced into Eq.(\ref{Rydb}), but in this paper we do not discuss this 
problem. Note only that the numerical values of the quantum defect $\Delta_{\ell}$ are uniformly related to the short-range (or non-Coulomb) component of the phase
shifts of elastic scattering of single-electron scattering by the two-electron positively charged ions \cite{Burke}. This directly follows from the unitarity of the
$S-$matrix (see, e.g., \cite{Peier}, \cite{Baz}).       

By using our computational results for the large number of bound states in the three-electron Li atom and in analogous Be$^{+}$, C$^{3+}$ and F$^{6+}$ ions (see Table 
IV) we were able to draw the energy levels of all computed doublet (bound) states in these atomic systems as functions of angular momentum $L$ of these states (see, Figs. 
1 - 4). Note that the total energies of all states shown on these Figures are lower than the corresponding threshold energies for these systems. We were able to calculate 
states with angular momentum L=20, see Table IV, whose energies are slightly lower than the estimated ionization limits. These threshold energies
(or ionization limit) coincide with the total, non-relativistic energies of the two-electron ions:
$E_{{}^{\infty}{\rm Li}^{+}}$ $\approx$ -7.279913 412669 305964 91895(15) $a.u.$, 
$E_{{}^{\infty}{\rm Be}^{2+}}$ $\approx$ -13.65556 623842 358670 208085(55) $a.u.$, 
$E_{{}^{\infty}{\rm C}^{4+}}$ $\approx$ -32.40624 660189 853031 055785(45) $a.u.$ and 
$E_{{}^{\infty}{\rm F}^{7+}}$ $\approx$ -75.53171 236395 949115(3) $a.u.$
In old books on atomic spectroscopy such pictures $E(L)$ (or diagrams) were called `spectral diagrams'. In the general case, the spectral diagram also depends upon the
total electron spin of the atom/ion, i.e. $E = E(L,S)$. Spectral diagrams are very useful tools to study various effects related with the electron density distribution in 
different bound $LS$-states of the atomic systems which contain the same number of electrons. Note that such spectral diagrams can be drawn for the neutral atoms as well as 
for various positively charged ions, e.g., for the Be atom and B$^{2+}$, N$^{4+}$ and other similar three-electron ions. The bound state spectra of the negatively charged 
ions, e.g., the Li$^{-}$ ion) contain only a very few bound states (usually one bound state \cite{Fro99}) and the corresponding spectral diagrams are very simple and not 
informative.

Let us discuss the spectral diagrams of the three-electron atomic systems shown on Figs. 1 - 4. As follows from these pictures the increase of the nuclear charge $Q$ in
these systems leads to the `hydrogenization' of the optical spectrum along the line: Li $\rightarrow$ Be$^{+}$ $\rightarrow$ C$^{3+}$ $\rightarrow$ F$^{6+}$. 
The energy levels are re-grouping (when $Q$ increases) into clusters which contain the energy levels with the same principal quantum number $n$. In other words, the 
differences between energies of levels with the same principal quantum number $n$ become much smaller than analogous differences between two energy levels with 
different principal quantum numbers $n$ and $n^{\prime}$. As one can see from our pictures such a clusterization rule is applied even to the energy levels with $n = 2$ 
and $n = 3$. Based on this observation we can predict that in the limit $Q \rightarrow \infty$ the bound state spectrum of three-electron ions looks like the bound state 
spectrum (or optical spectrum, for short) of a typical hydrogen-like ion in which, however, the ground state (or $1S^2-$state) is missing. Briefly, we can say that the 
`optical' spectrum of the doublet bound states in three-electron ions converges (at $Q \rightarrow \infty$) to the doublet spectrum of a hydrogen-like atomic system, 
where the ground $1s^2S-$state is missing. This explains why the traditional classification scheme used for bound state spectra in atomic spectroscopy is correct. For all 
atoms from the second row the ground state(s) must have the fundamental quantum number $n = 2$ (not $n = 1$, or $n = 3$). It is clear that changes in
the optical spectra of the three-electron atomic systems: Li, Be$^{+}$, C$^{3+}$ and F$^{6+}$ are directly related with the $Q-$dependent balance between the
electron-nucleus attractions and electron-electron repulsions. Another observation when comparing the spectral diagrams of the three-electron systems is the larger
relative stabilization or contraction of the 2$^2P$ level when growing the nuclear charge, while other low-lying levels are only slightly contracted.
There are many other observations which follow from Figs. 1 - 4, which shall stimulate future research.

Similarity between spectra of bound doublet ${}^{2}S-$states in different three-electron atomic systems can be seen from comparison of Figs.1 - 4 with each other. However, 
if we compare the same spectra reduced to the unit scale, then the observed agreement improves drastically. The procedure of reduction can be described as follows. First, 
for each three-electron atomic system with the nuclear charge $Q$ one needs to know the corresponding threshold energy $E_{tr}$. Usually, such an energy coincides with the
total energy of the ground $1^1S-$state of the two-electron ion which is formed from the incident three-electron atomic system after single-electron ionization. Second, 
the total energy of the ground $2^2S-$state $E_{gr}$ must be determined (i.e. measured and/or calculated) to the maximal numerical accuracy. The total energies of other
bound doublet states are recalculated with the use of the formula: $\tilde{E}_n = \frac{E_n - E_{tr}}{E_{gr} - E_{tr}}$. Finally, we have the new `energy' spectrum 
$\tilde{E}_{n}$. All eigenvalues from this spectrum are bounded between 0 and 1. Furthermore, there is only one limiting point for the spectrum of bound states $\tilde{E}_n$ 
and it coincides with 0 (or $E_{tr}$ in the original units). Briefly, we can say that all bound states $\tilde{E}_n$ can converge only to this `limiting' point. Since the 
threshold state is an actual state of any three-electron atomic system then the wave functions of bound states do not form a complete system of functions (for more detail,
see, e.g., \cite{Fro99}). The energy spectra $\tilde{E}_n$ of different three-electron atoms and ions (doublet states) once reduced to the same energy scale can be 
compared with each other directly.        

\section{Conclusion}

We have considered bound state spectra and properties of the doublet states in three-electron atomic systems. In this study we applied four different variational 
expansions for the bound state wave functions: (a) semi-exponential expansion in the relative coordinates, (b) expansion written in six-dimensional gaussoids, and (c) 
Hy-CI expansion for the low-lying $S$-, $P$-, and $D$-states of C$^{3+}$ and F$^{6+}$ ions, and (d) CI expansion for the $L \ge 3$ states. 
Very compact wave functions constructed with the use of the semi-exponential expansion in the relative coordinates allow one to determine a large number of
bound state properties, including the expectation values of electron-nucleus and electron-electron delta-functions and cusp values. The observed coincidence of the
computed cusp values with the predicted cusps (for Coulomb systems) is sufficient to recognize our trial wave functions accurate for numerical computations of all
bound state properties. Variational expansion in six-dimensional gaussoids is used to perform fast and accurate calculations of some excited states in three-electron 
atomic systems (atoms and ions). Results of such calculations include not only the total energies, but also a large number of other bound state properties. Moreover, 
numerical results of our calculations presented in Tables I and III allows us to correct a few mistakes/misprints made in earlier works in definition of these properties 
and/or in their numerical values. Our variational Hy-CI and CI wave function expansions are used to determine the total energies of various `rotationally' and `vibrationally' excited 
states in three-electron atoms and ions. In our calculations we consider a large number of the $S(L = 0)$-, $P(L = 1)$-, $D(L = 2)$-, $F(L = 3)$-, $G(L = 4)$-, $H(L = 5)$-, $I(L = 6)$-, $K(L = 7)$-, $L(L = 8)$-, $M(L = 9)$-, and $N(L = 10)$-states. 
Accurate results for highly excited rotational states have been determined in this study for the first time. The coincidence of our theoretically predicted spectrum (or computational spectrum) with the known optical spectrum of the three-electron Li atom is absolute \cite{Kik}, \cite{NIST} since we correctly 
predicted the actual order of different energy levels (or bound states) in this spectrum and evaluated the energy distances between different levels (to very good 
numerical accuracy). The order of the duplet energy levels in three-electron systems is $S <   P <   D <   F <   G <   H <   I <   K <   L <   M <   N <   \dots $ 

Our main goal in this study was to consider the bound state spectra and properties of the doublet states in three-electron atomic systems: Li atom, Be$^{+}$, C$^{3+}$
and F$^{6+}$ ions. The general structure of the bound state spectra in these three-electron atomic systems has been determined from the results of accurate numerical
computations performed with the use of the Hy-CI and CI variational expansions. By varying the nuclear electric charge $Q$ we investigated changes in the bound state 
spectra of such systems. The overall and partial contributions of the electron-electron correlations in the total energies of bound (doublet) states have been 
evaluated to high numerical accuracy. Formally, it is the first theoretical study in which a large number of bound states in a few three-electron atoms/ions are 
determined in highly accurate computations. In contrast with many modern studies we considered not one, or two bound states in three-electron atoms/ions, but essentially 
a whole bound state spectrum for each of the systems mentioned in this study. At the post Hartree-Fock level of accuracy for the doublet (bound) states in three-electron 
atoms and ions this was made for the first time. In the future, we are planning to study some other aspects of physics of three-electron atomic systems.

\newpage

% PIC1  LI  ATOM 

\begin{sidewaysfigure}

\begin{center}
  \sansmath
  \begin{tikzpicture}[
    font=\sffamily,
    level/.style={black,thick},
    ionization/.style={black,dashed},scale=1.0
  ]
  \coordinate (sublevel) at (0, 8pt);

  \node at (-0.25,10.0) {n};
  \node at (0.5, 10.5){$^2$S};
  \node at (2.5, 10.5){$^2$P}; 
  \node at (4.5, 10.5){$^2$D};
  \node at (6.5, 10.5){$^2$F}; 
  \node at (8.5, 10.5){$^2$G};
  \node at (10.5, 10.5){$^2$H};
  \node at (12.5, 10.5){$^2$I}; 
  \node at (14.5, 10.5){$^2$K};
  \node at (16.5, 10.5){$^2$L};
  \node at (18.5, 10.5){$^2$M};
  \node at (20.5, 10.5){$^2$N};

  % S levels
  \node at (0.5,-0.20) {\scriptsize $2s$};
  \node at (0.5, 6.12) {\scriptsize $3s$};
  \node at (0.5, 7.91) {\scriptsize $4s$};
  \node at (0.5, 8.66) {\scriptsize $5s$};
%  \node at (0.5, 9.00) {\scriptsize $6s$};
% \node at (0.5, 9.08) {\scriptsize $7s$};
%  \node at (0.5, 9.26) {\scriptsize $8s$};

  \coordinate (S00) at (0, 0.00);
  \coordinate (S10) at (0, 6.32);
  \coordinate (S20) at (0, 8.11);
  \coordinate (S30) at (0, 8.86);
  \coordinate (S40) at (0, 9.20);
  \coordinate (S50) at (0, 9.45);

  % Draw main levels
   \foreach \level/\text in { 00/2,  10/3,  20/4,  30/5, 40/6, 50/7}
     \draw[level] (S\level) node[left=20pt] {} node[left]
      {\scriptsize {\text}} -- +(1.0, 0);

  % P levels          

  \node at (2.5,3.33){\scriptsize $2p$};
  \node at (2.5,6.96){\scriptsize $3p$};
  \node at (2.5,8.21){\scriptsize $4p$};
  \node at (2.5,8.80){\scriptsize $5p$};
%  \node at (2.5,9.10) {\scriptsize $6p$};
%  \node at (2.5,9.11) {\scriptsize $7p$};
%  \node at (2.5,9.42) {\scriptsize $8p$};

  \coordinate (P00) at (2, 3.53);
  \coordinate (P10) at (2, 7.16);
  \coordinate (P20) at (2, 8.41);
  \coordinate (P30) at (2, 9.00);
  \coordinate (P40) at (2, 9.30);
  \coordinate (P50) at (2, 9.50);

  % Draw main levels
  \foreach \level/\text in {00/2, 10/3, 20/4, 30/5, 40/6, 50/7}
    \draw[level] (P\level) node[left=20pt] {} node[left] 
    {\scriptsize {\text}} -- +(1.0, 0);

  % D levels          
  \node at (4.5,7.06){\scriptsize $3d$};
  \node at (4.5,8.26){\scriptsize $4d$};
  \node at (4.5,8.85){\scriptsize $5d$};
%  \node at (4.5,9.10) {\scriptsize $6d$};
%  \node at (4.5,9.62) {\scriptsize $7d$};
%  \node at (4.5,9.72) {\scriptsize $8d$};

  \coordinate (D00) at (4, 7.26);
  \coordinate (D10) at (4, 8.46);
  \coordinate (D20) at (4, 9.05);
  \coordinate (D30) at (4, 9.30);
  \coordinate (D40) at (4, 9.50);

  % Draw main levels
  \foreach \level/\text in {00/3, 10/4, 20/5, 30/6, 40/7}
    \draw[level] (D\level) node[left=20pt] {} node[left]
    {\scriptsize {\text}} -- +(1.0, 0);

  % F levels          
  \node at (6.5,8.31){\scriptsize $4f$};
  \node at (6.5,8.80){\scriptsize $5f$};
  \node at (6.5,9.15){\scriptsize $6f$};
%  \node at (6.5,9.12) {\scriptsize $7f$};
%  \node at (6.5,9.28) {\scriptsize $8f$};

  \coordinate (F00) at (6, 8.51);
  \coordinate (F10) at (6, 9.00);
  \coordinate (F20) at (6, 9.35);
  \coordinate (F30) at (6, 9.55);

  % Draw main levels
  \foreach \level/\text in {00/4, 10/5, 20/6, 30/7} 
    \draw[level] (F\level) node[left=20pt] {} node[left]
    {\scriptsize {\text}} -- +(1.0, 0);

  % G levels          
  \node at (8.5,8.80){\scriptsize $5g$};
  \node at (8.5,9.13){\scriptsize $6g$};
%  \node at (8.5,9.21) {\scriptsize $7g$};
%  \node at (8.5,9.28) {\scriptsize $8g$};
  \coordinate (G00) at (8,9.00);
  \coordinate (G10) at (8,9.33);  
  \coordinate (G20) at (8,9.51);

  % Draw main levels
  \foreach \level/\text in {00/5, 10/6, 20/7}
    \draw[level] (G\level) node[left=20pt] {} node[left]
     {\scriptsize {\text}} -- +(1.0, 0);

  % H levels
  \node at (10.5,9.12){\scriptsize $6h$};
%  \node at (10.5,9.11) {\scriptsize $7h$};
%  \node at (10.5,9.28) {\scriptsize $8h$};
  \coordinate (H00) at (10,9.32);
  \coordinate (H10) at (10,9.51);

  % Draw main levels
  \foreach \level/\text in {00/6, 10/7}
    \draw[level] (H\level) node[left=20pt] {} node[left]
   {\scriptsize {\text}} -- +(1.0, 0);

  % I levels
  \node at (12.5,9.31) {\scriptsize $7i$};
%  \node at (12.5,9.28) {\scriptsize $8i$};
  \coordinate (I00) at (12,9.51);

  % Draw main levels
  \foreach \level/\text in {00/7}
    \draw[level] (I\level) node[left=20pt] {} node[left]
    {\scriptsize {\text}} -- +(1.0, 0);

  % K levels

  \node at (14.5,9.43){\scriptsize $8k$};
  \coordinate (K00) at (14,9.63);

  % Draw main levels
  \foreach \level/\text in {00/8}
    \draw[level] (K\level) node[left=20pt] {} node[left]
%     {\footnotesize {\text}} -- +(1.0, 0);
   {\scriptsize   {\text}} -- +(1.0, 0);

  % L levels

  \node at (16.5,9.45){\scriptsize $9l$};
  \coordinate (L00) at (16,9.65);

  % Draw main levels
  \foreach \level/\text in {00/9}
    \draw[level] (L\level) node[left=20pt] {} node[left]
%     {\footnotesize {\text}} -- +(1.0, 0);
    {\scriptsize   {\text}} -- +(1.0, 0);

  % M levels

  \node at (18.5,9.47){\scriptsize $10m$};
  \coordinate (M00) at (18,9.67);

  % Draw main levels
  \foreach \level/\text in {00/10}
    \draw[level] (M\level) node[left=20pt] {} node[left]
%     {\footnotesize {\text}} -- +(1.0, 0);
   {\scriptsize   {\text}} -- +(1.0, 0);

  % N levels

  \node at (20.5,9.49){\scriptsize $11n$};
  \coordinate (N00) at (20,9.69);

  % Draw main levels
  \foreach \level/\text in {00/11}
    \draw[level] (N\level) node[left=20pt] {} node[left]
%     {\footnotesize {\text}} -- +(1.0, 0);
     {\scriptsize   {\text}} -- +(1.0, 0);

  % Ionization level
  \draw[ionization] (0, 10.0) node[left=20pt] {E(${\rm Li}^+$)}-- +( 21.0, 0);

  % Rydberg levels
  \draw[level] (0,9.8) node[left=20pt] {}-- +(1.0, 0);
  \draw[level] (0,9.9) node[left=20pt] {}-- +(1.0, 0); 

  \draw[level] (2,9.8) node[left=20pt] {}-- +(1.0, 0);
  \draw[level] (2,9.9) node[left=20pt] {}-- +(1.0, 0);

  \draw[level] (4,9.8) node[left=20pt] {}-- +(1.0, 0);
  \draw[level] (4,9.9) node[left=20pt] {}-- +(1.0, 0);

  \draw[level] (6,9.8) node[left=20pt] {}-- +(1.0, 0);
  \draw[level] (6,9.9) node[left=20pt] {}-- +(1.0, 0);

  \draw[level] (8,9.8) node[left=20pt] {}-- +(1.0, 0);
  \draw[level] (8,9.9) node[left=20pt] {}-- +(1.0, 0);

  \draw[level] (10,9.8) node[left=20pt] {}-- +(1.0, 0);
  \draw[level] (10,9.9) node[left=20pt] {}-- +(1.0, 0);

  \draw[level] (12,9.8) node[left=20pt] {}-- +(1.0, 0);
  \draw[level] (12,9.9) node[left=20pt] {}-- +(1.0, 0);

  \draw[level] (14,9.8) node[left=20pt] {}-- +(1.0, 0);
  \draw[level] (14,9.9) node[left=20pt] {}-- +(1.0, 0);

  \draw[level] (16,9.8) node[left=20pt] {}-- +(1.0, 0);
  \draw[level] (16,9.9) node[left=20pt] {}-- +(1.0, 0);

  \draw[level] (18,9.8) node[left=20pt] {}-- +(1.0, 0);
  \draw[level] (18,9.9) node[left=20pt] {}-- +(1.0, 0);

  \draw[level] (20,9.8) node[left=20pt] {}-- +(1.0, 0);
  \draw[level] (20,9.9) node[left=20pt] {}-- +(1.0, 0);

  \end{tikzpicture}
\end{center}

  \vspace{1cm}

  \captionof{figure}{Energy levels for the doublet states of the lithium atom. The threshold energy -7.279913 412669 \dots a.u. 
coincides with the total energy of the ground $1^1$S state of the two-electron Li$^{+}$ ion.}

\end{sidewaysfigure}
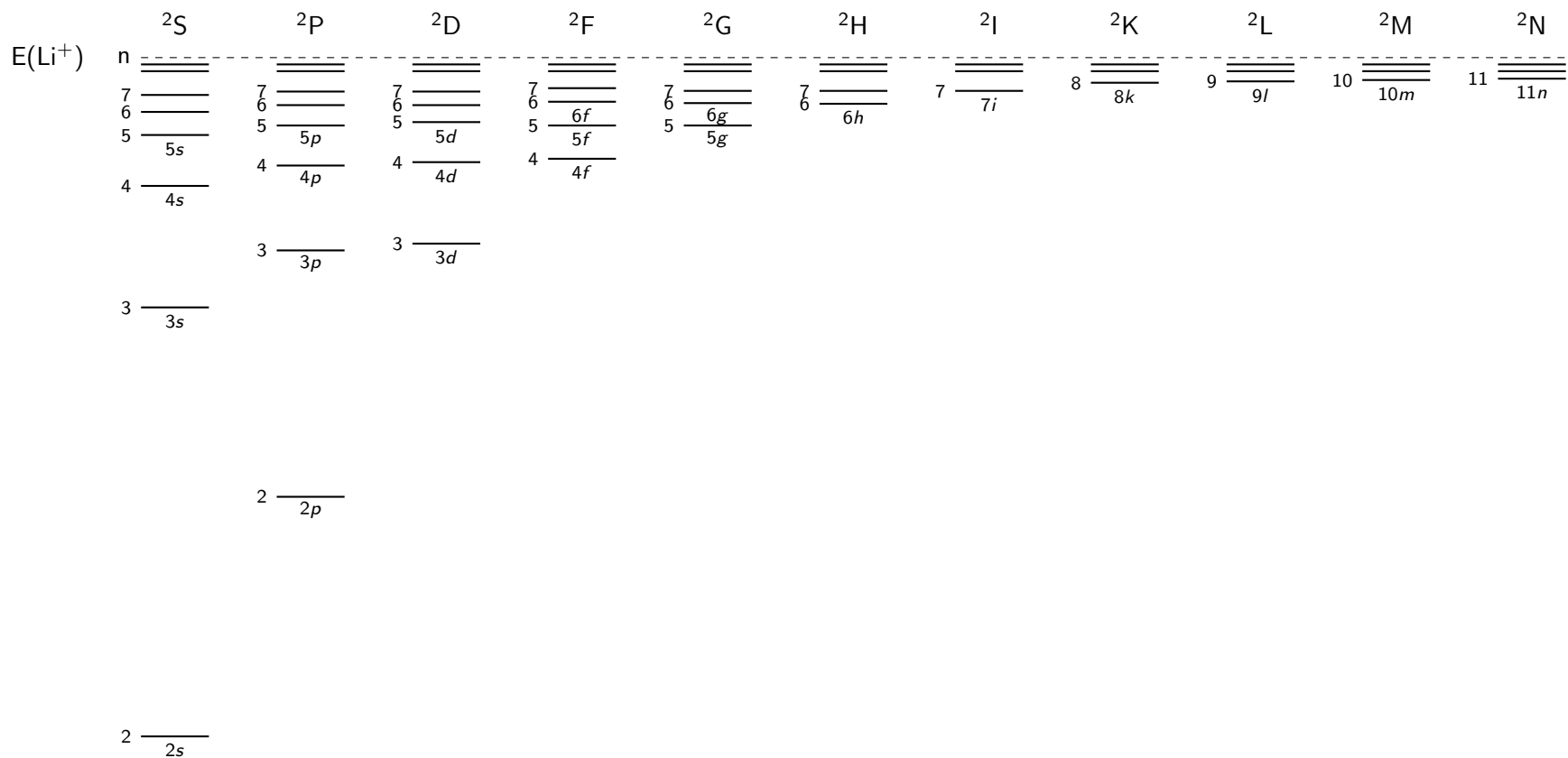

\newpage

% PIC2  Be+  ION 

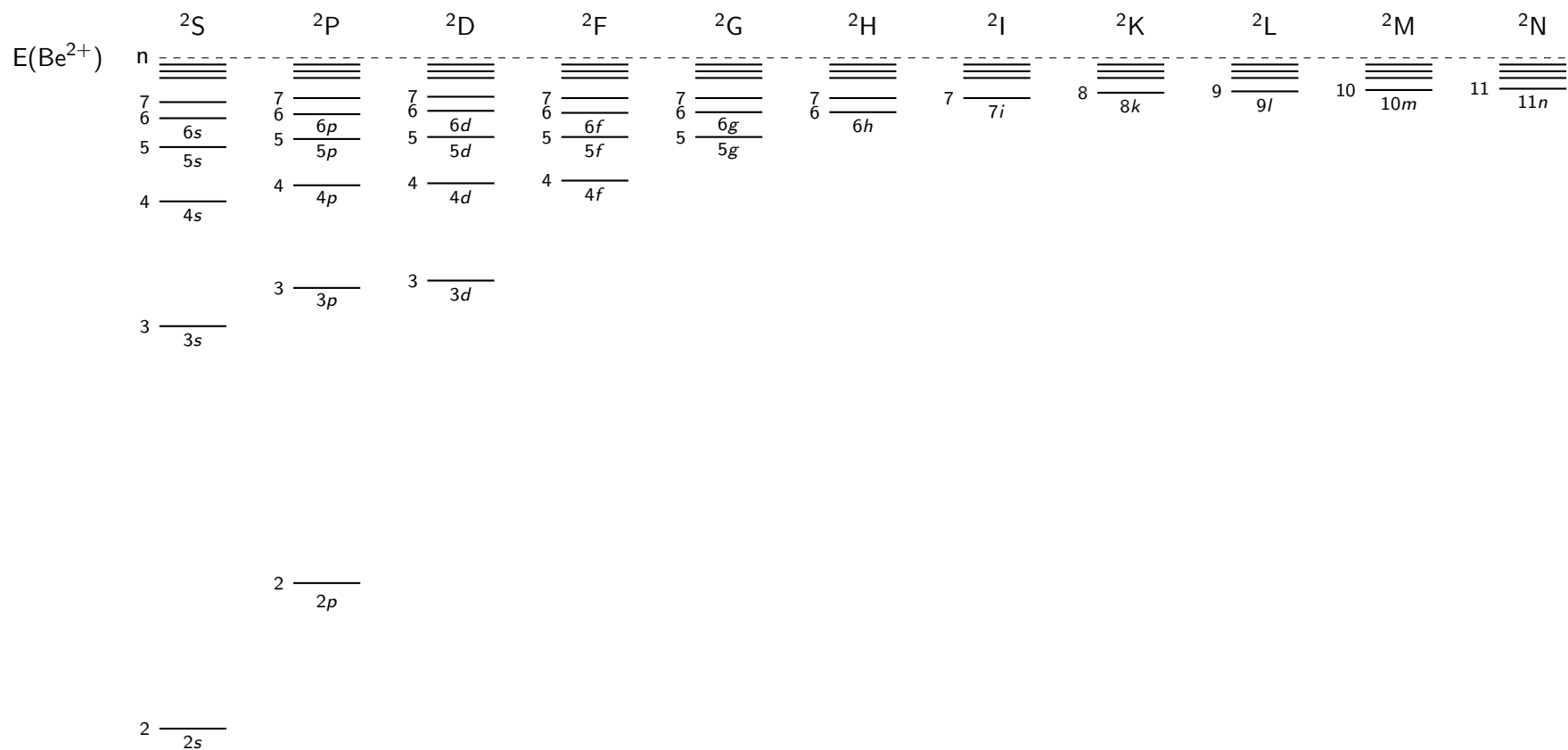
\begin{sidewaysfigure}

\begin{center}
  \sansmath
  \begin{tikzpicture}[
    font=\sffamily,
    level/.style={black,thick},
    ionization/.style={black,dashed},
  ]
  \coordinate (sublevel) at (0, 8pt);

  \node at (-0.25,10.0) {n};
  \node at (0.5, 10.5) {$^2$S};
  \node at (2.5, 10.5) {$^2$P};
  \node at (4.5, 10.5) {$^2$D};
  \node at (6.5, 10.5) {$^2$F};
  \node at (8.5, 10.5) {$^2$G};
  \node at (10.5, 10.5) {$^2$H};
  \node at (12.5, 10.5) {$^2$I};
  \node at (14.5, 10.5) {$^2$K};
  \node at (16.5, 10.5){$^2$L};
  \node at (18.5, 10.5){$^2$M};
  \node at (20.5, 10.5){$^2$N};

  % S levels
  \node at (0.5,-0.20) {\scriptsize $2s$};
  \node at (0.5, 5.80) {\scriptsize $3s$};
  \node at (0.5, 7.66) {\scriptsize $4s$};
  \node at (0.5, 8.47) {\scriptsize $5s$};
  \node at (0.5, 8.90) {\scriptsize $6s$};
% \node at (0.5, 9.08) {\scriptsize $7s$};
%  \node at (0.5, 9.26) {\scriptsize $8s$};

  \coordinate (S00) at (0, 0.00);
  \coordinate (S10) at (0, 6.00);
  \coordinate (S20) at (0, 7.86);
  \coordinate (S30) at (0, 8.67);
  \coordinate (S40) at (0, 9.10);
  \coordinate (S50) at (0, 9.34);

  % Draw main levels
   \foreach \level/\text in { 00/2,  10/3,  20/4,  30/5, 40/6, 50/7}
     \draw[level] (S\level) node[left=20pt] {} node[left]  
        {\scriptsize {\text}} -- +(1.0, 0);

  % P levels          
  \node at (2.5,1.87) {\scriptsize $2p$};
  \node at (2.5,6.37) {\scriptsize $3p$};
  \node at (2.5,7.90) {\scriptsize $4p$};
  \node at (2.5,8.59) {\scriptsize $5p$};
  \node at (2.5,8.96) {\scriptsize $6p$};
%  \node at (2.5,9.11) {\scriptsize $7p$};
%  \node at (2.5,9.42) {\scriptsize $8p$};

  \coordinate (P00) at (2, 2.17);
  \coordinate (P10) at (2, 6.57);
  \coordinate (P20) at (2, 8.10);
  \coordinate (P30) at (2, 8.79);
  \coordinate (P40) at (2, 9.16);
  \coordinate (P50) at (2, 9.40);

  % Draw main levels
  \foreach \level/\text in {00/2, 10/3, 20/4, 30/5, 40/6, 50/7}
    \draw[level] (P\level) node[left=20pt] {} node[left] 
     {\scriptsize {\text}} -- +(1.0, 0);

  % D levels          

  \node at (4.5,6.48) {\scriptsize $3d$};
  \node at (4.5,7.93) {\scriptsize $4d$};
  \node at (4.5,8.62) {\scriptsize $5d$};
  \node at (4.5,9.01) {\scriptsize $6d$};
%  \node at (4.5,9.62) {\scriptsize $7d$};
%  \node at (4.5,9.72) {\scriptsize $8d$};

  \coordinate (D00) at (4, 6.68);
  \coordinate (D10) at (4, 8.13);
  \coordinate (D20) at (4, 8.82);
  \coordinate (D30) at (4, 9.21);
  \coordinate (D40) at (4, 9.42);

  % Draw main levels
  \foreach \level/\text in {00/3, 10/4, 20/5, 30/6, 40/7}
    \draw[level] (D\level) node[left=20pt] {} node[left]
     {\scriptsize {\text}} -- +(1.0, 0);

  % F levels          
  \node at (6.5,7.97) {\scriptsize $4f$};
  \node at (6.5,8.62) {\scriptsize $5f$};
  \node at (6.5,8.98) {\scriptsize $6f$};
%  \node at (6.5,9.12) {\scriptsize $7f$};
%  \node at (6.5,9.28) {\scriptsize $8f$};

  \coordinate (F00) at (6, 8.17);
  \coordinate (F10) at (6, 8.82);
  \coordinate (F20) at (6, 9.18);
  \coordinate (F30) at (6, 9.40);

  % Draw main levels
  \foreach \level/\text in {00/4, 10/5, 20/6, 30/7} 
    \draw[level] (F\level) node[left=20pt] {} node[left]
     {\scriptsize {\text}} -- +(1.0, 0);

  % G levels          
  \node at (8.5,8.62) {\scriptsize $5g$};
  \node at (8.5,8.99) {\scriptsize $6g$};
%  \node at (8.5,9.21) {\scriptsize $7g$};
%  \node at (8.5,9.28) {\scriptsize $8g$};

  \coordinate (G00) at (8,8.82);
  \coordinate (G10) at (8,9.19);  
  \coordinate (G20) at (8,9.40);

  % Draw main levels
  \foreach \level/\text in {00/5, 10/6, 20/7}
    \draw[level] (G\level) node[left=20pt] {} node[left]
     {\scriptsize {\text}} -- +(1.0, 0);

  % H levels
  \node at (10.5,8.99) {\scriptsize $6h$};
%  \node at (10.5,9.11) {\scriptsize $7h$};
%  \node at (10.5,9.28) {\scriptsize $8h$};

  \coordinate (H00) at (10,9.19);
  \coordinate (H10) at (10,9.40);

  % Draw main levels
  \foreach \level/\text in {00/6, 10/7}
    \draw[level] (H\level) node[left=20pt] {} node[left]
     {\scriptsize {\text}} -- +(1.0, 0);

  % I levels
  \node at (12.5,9.20) {\scriptsize $7i$};
%  \node at (12.5,9.28) {\scriptsize $8i$};

  \coordinate (I00) at (12,9.40);

  % Draw main levels
  \foreach \level/\text in {00/7}
    \draw[level] (I\level) node[left=20pt] {} node[left]
     {\scriptsize {\text}} -- +(1.0, 0);

  % K levels

  \node at (14.5,9.28) {\scriptsize $8k$};
  \coordinate (K00) at (14,9.48);

  % Draw main levels
  \foreach \level/\text in {00/8}
    \draw[level] (K\level) node[left=20pt] {} node[left]
%     {\footnotesize {\text}} -- +(1.0, 0);
     {\scriptsize   {\text}} -- +(1.0, 0);

  % L levels

  \node at (16.5,9.30){\scriptsize $9l$};
  \coordinate (L00) at (16,9.50);

  % Draw main levels
  \foreach \level/\text in {00/9}
    \draw[level] (L\level) node[left=20pt] {} node[left]
%     {\footnotesize {\text}} -- +(1.0, 0);
    {\scriptsize   {\text}} -- +(1.0, 0);

  % M levels

  \node at (18.5,9.32){\scriptsize $10m$};
  \coordinate (M00) at (18,9.52);

  % Draw main levels
  \foreach \level/\text in {00/10}
    \draw[level] (M\level) node[left=20pt] {} node[left]
%     {\footnotesize {\text}} -- +(1.0, 0);
   {\scriptsize   {\text}} -- +(1.0, 0);

  % N levels

  \node at (20.5,9.34){\scriptsize $11n$};
  \coordinate (N00) at (20,9.54);

  % Draw main levels
  \foreach \level/\text in {00/11}
    \draw[level] (N\level) node[left=20pt] {} node[left]
%     {\footnotesize {\text}} -- +(1.0, 0);
     {\scriptsize   {\text}} -- +(1.0, 0);

  % Ionization level
  \draw[ionization] (0, 10.0) node[left=20pt] {E(${\rm Be}^{2+}$)}-- +( 21.0, 0);

  % Rydberg levels
  
  \draw[level] (0,9.7) node[left=20pt] {}-- +(1.0, 0); 
  \draw[level] (0,9.8) node[left=20pt] {}-- +(1.0, 0);
  \draw[level] (0,9.9) node[left=20pt] {}-- +(1.0, 0); 

  \draw[level] (2,9.7) node[left=20pt] {}-- +(1.0, 0);
  \draw[level] (2,9.8) node[left=20pt] {}-- +(1.0, 0);
  \draw[level] (2,9.9) node[left=20pt] {}-- +(1.0, 0);

  \draw[level] (4,9.7) node[left=20pt] {}-- +(1.0, 0);
  \draw[level] (4,9.8) node[left=20pt] {}-- +(1.0, 0);
  \draw[level] (4,9.9) node[left=20pt] {}-- +(1.0, 0);

  \draw[level] (6,9.7) node[left=20pt] {}-- +(1.0, 0);
  \draw[level] (6,9.8) node[left=20pt] {}-- +(1.0, 0);
  \draw[level] (6,9.9) node[left=20pt] {}-- +(1.0, 0);

  \draw[level] (8,9.7) node[left=20pt] {}-- +(1.0, 0);
  \draw[level] (8,9.8) node[left=20pt] {}-- +(1.0, 0);
  \draw[level] (8,9.9) node[left=20pt] {}-- +(1.0, 0);

  \draw[level] (10,9.7) node[left=20pt] {}-- +(1.0, 0);
  \draw[level] (10,9.8) node[left=20pt] {}-- +(1.0, 0);
  \draw[level] (10,9.9) node[left=20pt] {}-- +(1.0, 0);

  \draw[level] (12,9.7) node[left=20pt] {}-- +(1.0, 0);
  \draw[level] (12,9.8) node[left=20pt] {}-- +(1.0, 0);
  \draw[level] (12,9.9) node[left=20pt] {}-- +(1.0, 0);

  \draw[level] (14,9.7) node[left=20pt] {}-- +(1.0, 0);
  \draw[level] (14,9.8) node[left=20pt] {}-- +(1.0, 0);
  \draw[level] (14,9.9) node[left=20pt] {}-- +(1.0, 0);

  \draw[level] (16,9.7) node[left=20pt] {}-- +(1.0, 0);
  \draw[level] (16,9.8) node[left=20pt] {}-- +(1.0, 0);
  \draw[level] (16,9.9) node[left=20pt] {}-- +(1.0, 0);

  \draw[level] (18,9.7) node[left=20pt] {}-- +(1.0, 0);
  \draw[level] (18,9.8) node[left=20pt] {}-- +(1.0, 0);
  \draw[level] (18,9.9) node[left=20pt] {}-- +(1.0, 0);

  \draw[level] (20,9.7) node[left=20pt] {}-- +(1.0, 0);
  \draw[level] (20,9.8) node[left=20pt] {}-- +(1.0, 0);
  \draw[level] (20,9.9) node[left=20pt] {}-- +(1.0, 0);

  \end{tikzpicture}
\end{center}

\vspace{1cm}

\captionof{figure}{Energy levels for the doublet states of the Be$^+$ ion. The threshold 
energy -13.65556 623842 \dots a.u. coincides with the total energy of the ground $1^1$S state of the two-electron Be$^{2+}$ ion.}

\end{sidewaysfigure}

% PIC2 C3+  ION 

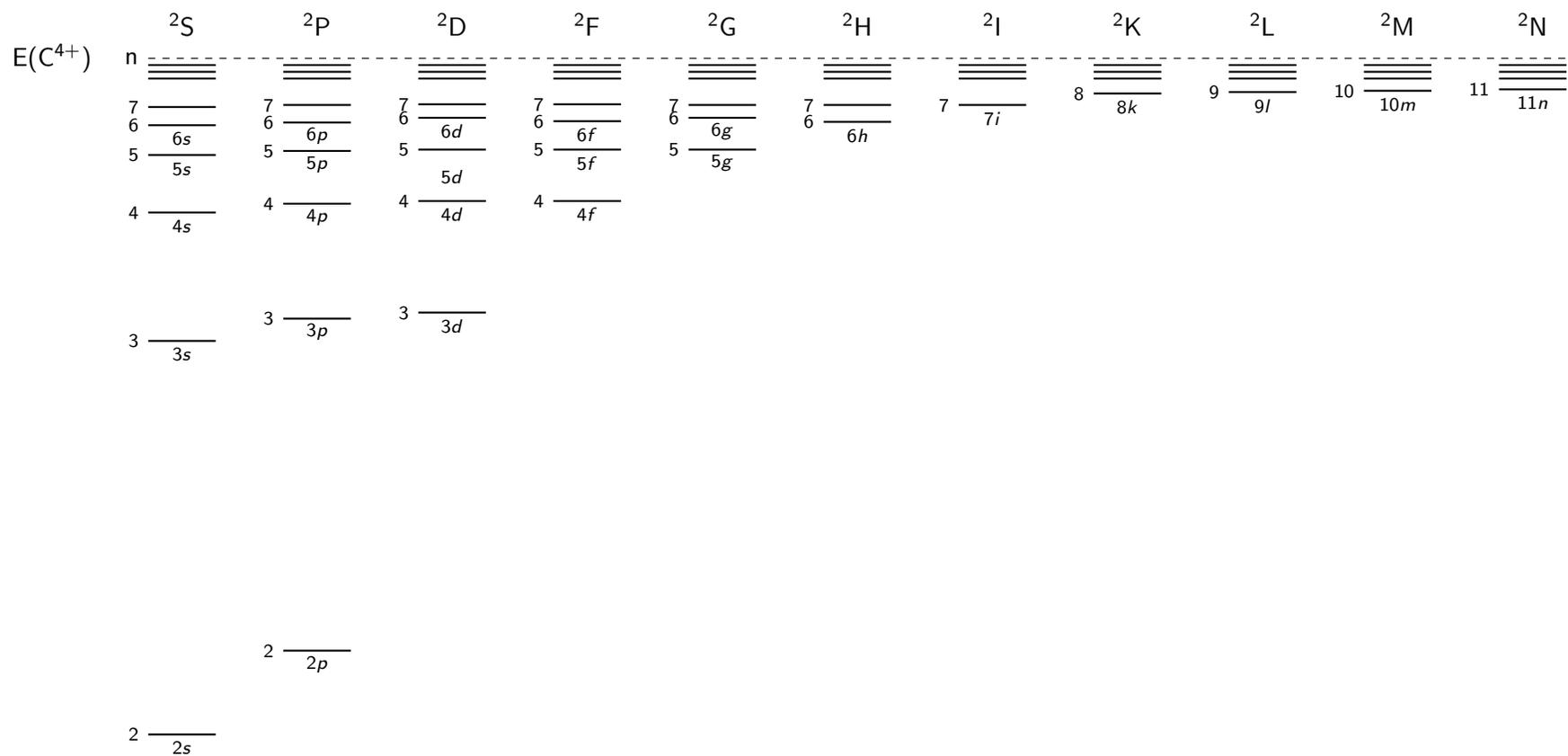
\begin{sidewaysfigure}

\begin{center}
  \sansmath
  \begin{tikzpicture}[
    font=\sffamily,
    level/.style={black,thick},
    ionization/.style={black,dashed},
  ]
  \coordinate (sublevel) at (0, 8pt);

  \node at (-0.25,10.0) {n};
  \node at (0.5, 10.5) {$^2$S};
  \node at (2.5, 10.5) {$^2$P};
  \node at (4.5, 10.5) {$^2$D};
  \node at (6.5, 10.5) {$^2$F};
  \node at (8.5, 10.5) {$^2$G};
  \node at (10.5, 10.5) {$^2$H};
  \node at (12.5, 10.5) {$^2$I};
  \node at (14.5, 10.5) {$^2$K};
  \node at (16.5, 10.5){$^2$L};
  \node at (18.5, 10.5){$^2$M};
  \node at (20.5, 10.5){$^2$N};

  % S levels

  \node at (0.5,-0.20) {\scriptsize $2s$};
  \node at (0.5, 5.62) {\scriptsize $3s$};
  \node at (0.5, 7.52) {\scriptsize $4s$};
  \node at (0.5, 8.37) {\scriptsize $5s$};
  \node at (0.5, 8.81) {\scriptsize $6s$};
% \node at (0.5, 9.08) {\scriptsize $7s$};
%  \node at (0.5, 9.26) {\scriptsize $8s$};

  \coordinate (S00) at (0, 0.00);
  \coordinate (S10) at (0, 5.82);
  \coordinate (S20) at (0, 7.72);
  \coordinate (S30) at (0, 8.57);
  \coordinate (S40) at (0, 9.01);
  \coordinate (S50) at (0, 9.28);
  \coordinate (S60) at (0, 9.46);

  % Draw main levels
   \foreach \level/\text in { 00/2,  10/3,  20/4,  30/5, 40/6, 50/7}
     \draw[level] (S\level) node[left=20pt] {} node[left]  
%        {\footnotesize {\text}} -- +(1.0, 0);
         {\scriptsize   {\text}} -- +(1.0, 0);

  % P levels          
 
  \node at (2.5,1.04) {\scriptsize $2p$};
  \node at (2.5,5.95) {\scriptsize $3p$};
  \node at (2.5,7.65) {\scriptsize $4p$};
  \node at (2.5,8.43) {\scriptsize $5p$};
  \node at (2.5,8.85) {\scriptsize $6p$};
%  \node at (2.5,9.11) {\scriptsize $7p$};
%  \node at (2.5,9.42) {\scriptsize $8p$};

  \coordinate (P00) at (2, 1.24);
  \coordinate (P10) at (2, 6.15);
  \coordinate (P20) at (2, 7.85);
  \coordinate (P30) at (2, 8.63);
  \coordinate (P40) at (2, 9.05);
  \coordinate (P50) at (2, 9.31);
%  \coordinate (P60) at (2, 9.60);

  % Draw main levels
  \foreach \level/\text in {00/2, 10/3, 20/4, 30/5, 40/6, 50/7}
    \draw[level] (P\level) node[left=20pt] {} node[left] 
%     {\footnotesize {\text}} -- +(1.0, 0);
      {\scriptsize   {\text}} -- +(1.0, 0);

  % D levels  

  \node at (4.5,6.04) {\scriptsize $3d$};
  \node at (4.5,7.69) {\scriptsize $4d$};
  \node at (4.5,8.25) {\scriptsize $5d$};
  \node at (4.5,8.92) {\scriptsize $6d$};
%  \node at (4.5,9.62) {\scriptsize $7d$};
%  \node at (4.5,9.72) {\scriptsize $8d$};

  \coordinate (D00) at (4, 6.24);
  \coordinate (D10) at (4, 7.89);
  \coordinate (D20) at (4, 8.65);
  \coordinate (D30) at (4, 9.12);
  \coordinate (D40) at (4, 9.32);
  \coordinate (D50) at (4, 9.62);

  % Draw main levels
  \foreach \level/\text in {00/3, 10/4, 20/5, 30/6, 40/7}
    \draw[level] (D\level) node[left=20pt] {} node[left]
%     {\footnotesize {\text}} -- +(1.0, 0);
      {\scriptsize   {\text}} -- +(1.0, 0);

  % F levels    

  \node at (6.5,7.69) {\scriptsize $4f$};
  \node at (6.5,8.45) {\scriptsize $5f$};
  \node at (6.5,8.87) {\scriptsize $6f$};
%  \node at (6.5,9.12) {\scriptsize $7f$};
%  \node at (6.5,9.28) {\scriptsize $8f$};

  \coordinate (F00) at (6, 7.89);
  \coordinate (F10) at (6, 8.65);
  \coordinate (F20) at (6, 9.07);
  \coordinate (F30) at (6, 9.32);
  \coordinate (F40) at (6, 9.48);

  % Draw main levels
  \foreach \level/\text in {00/4, 10/5, 20/6, 30/7} 
    \draw[level] (F\level) node[left=20pt] {} node[left]
%     {\footnotesize {\text}} -- +(1.0, 0);
      {\scriptsize   {\text}} -- +(1.0, 0);

  % G levels          

  \node at (8.5,8.45) {\scriptsize $5g$};
  \node at (8.5,8.92) {\scriptsize $6g$};
%  \node at (8.5,9.21) {\scriptsize $7g$};
%  \node at (8.5,9.28) {\scriptsize $8g$};

  \coordinate (G00) at (8,8.65);
  \coordinate (G10) at (8,9.12);  
  \coordinate (G20) at (8,9.31);
  \coordinate (G30) at (8,9.48);

  % Draw main levels
  \foreach \level/\text in {00/5, 10/6, 20/7}
    \draw[level] (G\level) node[left=20pt] {} node[left]
%     {\footnotesize {\text}} -- +(1.0, 0);
      {\scriptsize   {\text}} -- +(1.0, 0);

  % H levels

  \node at (10.5,8.86) {\scriptsize $6h$};
%  \node at (10.5,9.11) {\scriptsize $7h$};
%  \node at (10.5,9.28) {\scriptsize $8h$};
  \coordinate (H00) at (10,9.06);
  \coordinate (H10) at (10,9.31);
  \coordinate (H20) at (10,9.48);

  % Draw main levels
  \foreach \level/\text in {00/6, 10/7}
    \draw[level] (H\level) node[left=20pt] {} node[left]
%     {\footnotesize {\text}} -- +(1.0, 0);
      {\scriptsize   {\text}} -- +(1.0, 0);

  % I levels

  \node at (12.5,9.11) {\scriptsize $7i$};
%  \node at (12.5,9.28) {\scriptsize $8i$};
  \coordinate (I00) at (12,9.31);
  \coordinate (I10) at (12,9.48);

  % Draw main levels
  \foreach \level/\text in {00/7}
    \draw[level] (I\level) node[left=20pt] {} node[left]
%     {\footnotesize {\text}} -- +(1.0, 0);
      {\scriptsize   {\text}} -- +(1.0, 0);

  % K levels

  \node at (14.5,9.28) {\scriptsize $8k$};
  \coordinate (K00) at (14,9.48);

  % Draw main levels
  \foreach \level/\text in {00/8}
    \draw[level] (K\level) node[left=20pt] {} node[left]
%     {\footnotesize {\text}} -- +(1.0, 0);
     {\scriptsize   {\text}} -- +(1.0, 0);

  % L levels

  \node at (16.5,9.30){\scriptsize $9l$};
  \coordinate (L00) at (16,9.50) ;

  % Draw main levels
  \foreach \level/\text in {00/9}
    \draw[level] (L\level) node[left=20pt] {} node[left]
%     {\footnotesize {\text}} -- +(1.0, 0);
    {\scriptsize   {\text}} -- +(1.0, 0);

  % M levels

  \node at (18.5,9.32){\scriptsize $10m$};
  \coordinate (M00) at (18,9.52);

  % Draw main levels
  \foreach \level/\text in {00/10}
    \draw[level] (M\level) node[left=20pt] {} node[left]
%     {\footnotesize {\text}} -- +(1.0, 0);
   {\scriptsize   {\text}} -- +(1.0, 0);

  % N levels

  \node at (20.5,9.34){\scriptsize $11n$};
  \coordinate (N00) at (20,9.54);

  % Draw main levels
  \foreach \level/\text in {00/11}
    \draw[level] (N\level) node[left=20pt] {} node[left]
%     {\footnotesize {\text}} -- +(1.0, 0);
     {\scriptsize   {\text}} -- +(1.0, 0);

  % Ionization level
  \draw[ionization] (0, 10.0) node[left=20pt] {E(${\rm C}^{4+}$)}-- +( 21.0, 0);

  % Rydberg levels
 
  \draw[level] (0,9.7) node[left=20pt] {}-- +(1.0, 0); 
  \draw[level] (0,9.8) node[left=20pt] {}-- +(1.0, 0);
  \draw[level] (0,9.9) node[left=20pt] {}-- +(1.0, 0); 

  \draw[level] (2,9.7) node[left=20pt] {}-- +(1.0, 0);
  \draw[level] (2,9.8) node[left=20pt] {}-- +(1.0, 0);
  \draw[level] (2,9.9) node[left=20pt] {}-- +(1.0, 0);

  \draw[level] (4,9.7) node[left=20pt] {}-- +(1.0, 0);
  \draw[level] (4,9.8) node[left=20pt] {}-- +(1.0, 0);
  \draw[level] (4,9.9) node[left=20pt] {}-- +(1.0, 0);

  \draw[level] (6,9.7) node[left=20pt] {}-- +(1.0, 0);
  \draw[level] (6,9.8) node[left=20pt] {}-- +(1.0, 0);
  \draw[level] (6,9.9) node[left=20pt] {}-- +(1.0, 0);

  \draw[level] (8,9.7) node[left=20pt] {}-- +(1.0, 0);
  \draw[level] (8,9.8) node[left=20pt] {}-- +(1.0, 0);
  \draw[level] (8,9.9) node[left=20pt] {}-- +(1.0, 0);

  \draw[level] (10,9.7) node[left=20pt] {}-- +(1.0, 0);
  \draw[level] (10,9.8) node[left=20pt] {}-- +(1.0, 0);
  \draw[level] (10,9.9) node[left=20pt] {}-- +(1.0, 0);

  \draw[level] (12,9.7) node[left=20pt] {}-- +(1.0, 0);
  \draw[level] (12,9.8) node[left=20pt] {}-- +(1.0, 0);
  \draw[level] (12,9.9) node[left=20pt] {}-- +(1.0, 0);

  \draw[level] (14,9.7) node[left=20pt] {}-- +(1.0, 0);
  \draw[level] (14,9.8) node[left=20pt] {}-- +(1.0, 0);
  \draw[level] (14,9.9) node[left=20pt] {}-- +(1.0, 0);

  \draw[level] (16,9.7) node[left=20pt] {}-- +(1.0, 0);
  \draw[level] (16,9.8) node[left=20pt] {}-- +(1.0, 0);
  \draw[level] (16,9.9) node[left=20pt] {}-- +(1.0, 0);

  \draw[level] (18,9.7) node[left=20pt] {}-- +(1.0, 0);
  \draw[level] (18,9.8) node[left=20pt] {}-- +(1.0, 0);
  \draw[level] (18,9.9) node[left=20pt] {}-- +(1.0, 0);

  \draw[level] (20,9.7) node[left=20pt] {}-- +(1.0, 0);
  \draw[level] (20,9.8) node[left=20pt] {}-- +(1.0, 0);
  \draw[level] (20,9.9) node[left=20pt] {}-- +(1.0, 0);

  \end{tikzpicture}
 \vspace{1cm}
\end{center}

\captionof{figure}{Energy levels for the doublet states of the C$^{3+}$ ion. The threshold 
energy -32.40624 660189 \dots a.u. coincides with the total energy of the ground $1^1$S state of the two-electron C$^{4+}$ ion.} 

\end{sidewaysfigure}

\newpage

% PIC2  F6+  ION 

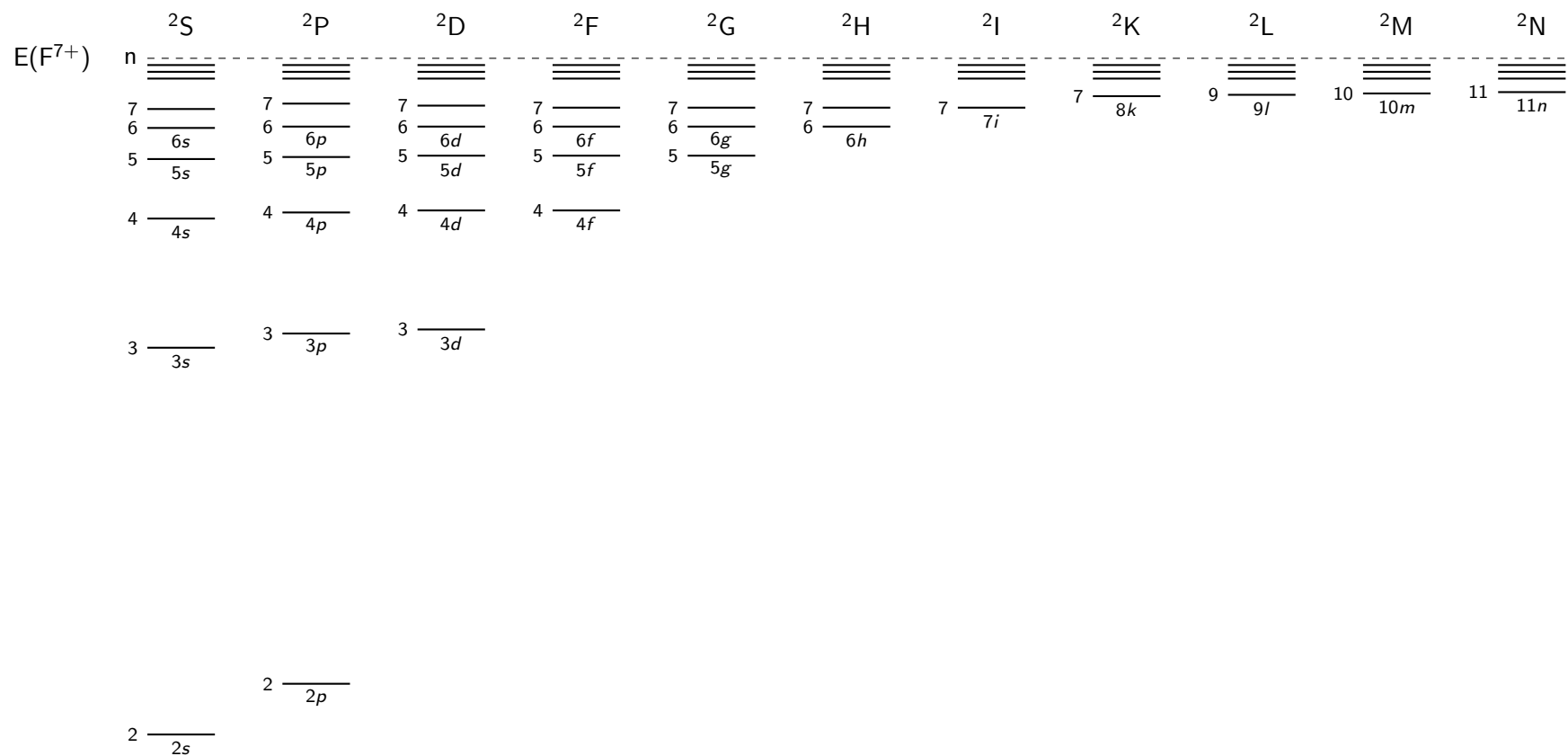
\begin{sidewaysfigure}

\begin{center}
  \sansmath
  \begin{tikzpicture}[
    font=\sffamily,
    level/.style={black,thick},
    ionization/.style={black,dashed},
  ]
  \coordinate (sublevel) at (0, 8pt);

  \node at (-0.25,10.0) {n};
  \node at (0.5, 10.5) {$^2$S};
  \node at (2.5, 10.5) {$^2$P};
  \node at (4.5, 10.5) {$^2$D};
  \node at (6.5, 10.5) {$^2$F};
  \node at (8.5, 10.5) {$^2$G};
  \node at (10.5, 10.5) {$^2$H};
  \node at (12.5, 10.5) {$^2$I};
  \node at (14.5, 10.5) {$^2$K};
  \node at (16.5, 10.5){$^2$L};
  \node at (18.5, 10.5){$^2$M};
  \node at (20.5, 10.5){$^2$N};

  % S levels

  \node at (0.5,-0.20) {\scriptsize $2s$};
  \node at (0.5, 5.52) {\scriptsize $3s$};
  \node at (0.5, 7.43) {\scriptsize $4s$};
  \node at (0.5, 8.31) {\scriptsize $5s$};
  \node at (0.5, 8.77) {\scriptsize $6s$};
% \node at (0.5, 9.05) {\scriptsize $7s$};
%  \node at (0.5, 9.23) {\scriptsize $8s$};

  \coordinate (S00) at (0, 0.00);
  \coordinate (S10) at (0, 5.72);
  \coordinate (S20) at (0, 7.63);
  \coordinate (S30) at (0, 8.51);
  \coordinate (S40) at (0, 8.97);
  \coordinate (S50) at (0, 9.25);
  \coordinate (S60) at (0, 9.43);

  % Draw main levels
   \foreach \level/\text in { 00/2,  10/3,  20/4,  30/5, 40/6, 50/7}
     \draw[level] (S\level) node[left=20pt] {} node[left]  
        {\scriptsize {\text}} -- +(1.0, 0);

  % P levels          

  \node at (2.5,0.55) {\scriptsize $2p$};
  \node at (2.5,5.73) {\scriptsize $3p$};
  \node at (2.5,7.52) {\scriptsize $4p$};
  \node at (2.5,8.34) {\scriptsize $5p$};
  \node at (2.5,8.79) {\scriptsize $6p$};
%  \node at (2.5,9.13) {\scriptsize $7p$};
%  \node at (2.5,9.23) {\scriptsize $8p$};

  \coordinate (P00) at (2, 0.75);
  \coordinate (P10) at (2, 5.93);
  \coordinate (P20) at (2, 7.72);
  \coordinate (P30) at (2, 8.54);
  \coordinate (P40) at (2, 8.99);
  \coordinate (P50) at (2, 9.33);
  \coordinate (P60) at (2, 9.43);

  % Draw main levels
  \foreach \level/\text in {00/2, 10/3, 20/4, 30/5, 40/6, 50/7}
    \draw[level] (P\level) node[left=20pt] {} node[left] 
     {\scriptsize {\text}} -- +(1.0, 0);

  % D levels         

  \node at (4.5,5.79) {\scriptsize $3d$};
  \node at (4.5,7.55) {\scriptsize $4d$};
  \node at (4.5,8.36) {\scriptsize $5d$};
  \node at (4.5,8.79) {\scriptsize $6d$};
%  \node at (4.5,9.10) {\scriptsize $7d$};
%  \node at (4.5,9.25) {\scriptsize $8d$};

  \coordinate (D00) at (4, 5.99);
  \coordinate (D10) at (4, 7.75);
  \coordinate (D20) at (4, 8.56);
  \coordinate (D30) at (4, 8.99);
  \coordinate (D40) at (4, 9.30);
  \coordinate (D50) at (4, 9.44);

  % Draw main levels
  \foreach \level/\text in {00/3, 10/4, 20/5, 30/6, 40/7}
    \draw[level] (D\level) node[left=20pt] {} node[left]
     {\scriptsize {\text}} -- +(1.0, 0);

  % F levels         

  \node at (6.5,7.55) {\scriptsize $4f$};
  \node at (6.5,8.36) {\scriptsize $5f$};
  \node at (6.5,8.79) {\scriptsize $6f$};
%  \node at (6.5,9.17) {\scriptsize $7f$};
%  \node at (6.5,9.24) {\scriptsize $8f$};

  \coordinate (F00) at (6, 7.75);
  \coordinate (F10) at (6, 8.56);
  \coordinate (F20) at (6, 8.99);
  \coordinate (F30) at (6, 9.27);
  \coordinate (F40) at (6, 9.44);

  % Draw main levels
  \foreach \level/\text in {00/4, 10/5, 20/6, 30/7} 
    \draw[level] (F\level) node[left=20pt] {} node[left]
     {\scriptsize {\text}} -- +(1.0, 0);

  % G levels         

  \node at (8.5,8.36) {\scriptsize $5g$};
  \node at (8.5,8.79) {\scriptsize $6g$};
%  \node at (8.5,9.22) {\scriptsize $7g$};
%  \node at (8.5,9.24) {\scriptsize $8g$};

  \coordinate (G00) at (8,8.56);
  \coordinate (G10) at (8,8.99);  
  \coordinate (G20) at (8,9.27);
  \coordinate (G30) at (8,9.44);

  % Draw main levels
  \foreach \level/\text in {00/5, 10/6, 20/7}
    \draw[level] (G\level) node[left=20pt] {} node[left]
     {\scriptsize {\text}} -- +(1.0, 0);

  % H levels

  \node at (10.5,8.79) {\scriptsize $6h$};
%  \node at (10.5,9.17) {\scriptsize $7h$};
%  \node at (10.5,9.24) {\scriptsize $8h$};

  \coordinate (H00) at (10,8.99);
  \coordinate (H10) at (10,9.27);
  \coordinate (H20) at (10,9.44);

  % Draw main levels
  \foreach \level/\text in {00/6, 10/7}
    \draw[level] (H\level) node[left=20pt] {} node[left]
     {\scriptsize {\text}} -- +(1.0, 0);

  % I levels
  \node at (12.5,9.07) {\scriptsize $7i$};
%  \node at (12.5,9.24) {\scriptsize $8i$};

  \coordinate (I00) at (12,9.27);
  \coordinate (I10) at (12,9.44);
  % Draw main levels
  \foreach \level/\text in {00/7}
    \draw[level] (I\level) node[left=20pt] {} node[left]
     {\scriptsize {\text}} -- +(1.0, 0);

  % K levels
  \node at (14.5,9.24) {\scriptsize $8k$};
  \coordinate (K00) at (14,9.44);
  % Draw main levels
  \foreach \level/\text in {00/7}
    \draw[level] (K\level) node[left=20pt] {} node[left]
     {\scriptsize {\text}} -- +(1.0, 0);

  % L levels

  \node at (16.5,9.26){\scriptsize $9l$};
  \coordinate (L00) at (16,9.46);

  % Draw main levels
  \foreach \level/\text in {00/9}
    \draw[level] (L\level) node[left=20pt] {} node[left]
%     {\footnotesize {\text}} -- +(1.0, 0);
    {\scriptsize   {\text}} -- +(1.0, 0);

  % M levels

  \node at (18.5,9.28){\scriptsize $10m$};
  \coordinate (M00) at (18,9.48);

  % Draw main levels
  \foreach \level/\text in {00/10}
    \draw[level] (M\level) node[left=20pt] {} node[left]
%     {\footnotesize {\text}} -- +(1.0, 0);
   {\scriptsize   {\text}} -- +(1.0, 0);

  % N levels

  \node at (20.5,9.30){\scriptsize $11n$};
  \coordinate (N00) at (20,9.50);

  % Draw main levels
  \foreach \level/\text in {00/11}
    \draw[level] (N\level) node[left=20pt] {} node[left]
%     {\footnotesize {\text}} -- +(1.0, 0);
     {\scriptsize   {\text}} -- +(1.0, 0);

  % Ionization level
  \draw[ionization] (0, 10.0) node[left=20pt] {E(${\rm F}^{7+}$)}-- +( 21.0, 0);

  % Rydberg levels
  
  \draw[level] (0,9.7) node[left=20pt] {}-- +(1.0, 0); 
  \draw[level] (0,9.8) node[left=20pt] {}-- +(1.0, 0);
  \draw[level] (0,9.9) node[left=20pt] {}-- +(1.0, 0); 

  \draw[level] (2,9.7) node[left=20pt] {}-- +(1.0, 0);
  \draw[level] (2,9.8) node[left=20pt] {}-- +(1.0, 0);
  \draw[level] (2,9.9) node[left=20pt] {}-- +(1.0, 0);

  \draw[level] (4,9.7) node[left=20pt] {}-- +(1.0, 0);
  \draw[level] (4,9.8) node[left=20pt] {}-- +(1.0, 0);
  \draw[level] (4,9.9) node[left=20pt] {}-- +(1.0, 0);

  \draw[level] (6,9.7) node[left=20pt] {}-- +(1.0, 0);
  \draw[level] (6,9.8) node[left=20pt] {}-- +(1.0, 0);
  \draw[level] (6,9.9) node[left=20pt] {}-- +(1.0, 0);

  \draw[level] (8,9.7) node[left=20pt] {}-- +(1.0, 0);
  \draw[level] (8,9.8) node[left=20pt] {}-- +(1.0, 0);
  \draw[level] (8,9.9) node[left=20pt] {}-- +(1.0, 0);

  \draw[level] (10,9.7) node[left=20pt] {}-- +(1.0, 0);
  \draw[level] (10,9.8) node[left=20pt] {}-- +(1.0, 0);
  \draw[level] (10,9.9) node[left=20pt] {}-- +(1.0, 0);

  \draw[level] (12,9.7) node[left=20pt] {}-- +(1.0, 0);
  \draw[level] (12,9.8) node[left=20pt] {}-- +(1.0, 0);
  \draw[level] (12,9.9) node[left=20pt] {}-- +(1.0, 0);

  \draw[level] (14,9.7) node[left=20pt] {}-- +(1.0, 0);
  \draw[level] (14,9.8) node[left=20pt] {}-- +(1.0, 0);
  \draw[level] (14,9.9) node[left=20pt] {}-- +(1.0, 0);

  \draw[level] (16,9.7) node[left=20pt] {}-- +(1.0, 0);
  \draw[level] (16,9.8) node[left=20pt] {}-- +(1.0, 0);
  \draw[level] (16,9.9) node[left=20pt] {}-- +(1.0, 0);

  \draw[level] (18,9.7) node[left=20pt] {}-- +(1.0, 0);
  \draw[level] (18,9.8) node[left=20pt] {}-- +(1.0, 0);
  \draw[level] (18,9.9) node[left=20pt] {}-- +(1.0, 0);

  \draw[level] (20,9.7) node[left=20pt] {}-- +(1.0, 0);
  \draw[level] (20,9.8) node[left=20pt] {}-- +(1.0, 0);
  \draw[level] (20,9.9) node[left=20pt] {}-- +(1.0, 0);

  \end{tikzpicture}
\vspace{1cm}

\end{center}

\captionof{figure}{Energy levels for the doublet states of the F$^{6+}$ ion. The threshold 
energy -75.53171 236395 \dots a.u. coincides with the total energy of the ground $1^1$S state of the two-electron F$^{7+}$ ion.} 

\end{sidewaysfigure}

\newpage

% TABLE I

 \begin{table}[tbp]
   \caption{Calculation of properties by using the semi-exponential variational expansion. 
            The expectation values of a number of electron-nuclear ($en$) and electron-electron ($ee$) 
            properties of the ground $2^2S-$states of the ${}^{\infty}$Li atom and in the ${}^{\infty}$Be$^{+}$
            and ${}^{\infty}$B$^{2+}$ ions (in atomic units). The notations $\langle V \rangle$ and $\langle T \rangle$
            stand for the expectation values of the potential and kinetic energy, respectively.}
     \begin{center}
%     \scalebox{0.72}{%
     \begin{tabular}{| c | c | c | c | c | c | c |}
       \hline\hline          
 atom/ion  & state  & $\langle r^{-1}_{eN} \rangle$ & $\langle r_{eN} \rangle$ & $\langle r^2_{eN} \rangle$  & $\langle r^3_{eN} \rangle$  &  $\langle r^4_{eN} \rangle$ \\
     \hline
 Li       & $2^2S$ & 1.906 035 791 & 1.663 195 075 & 6.118 405 34 & 30.869 167 & 183.374 94 \\

 Be$^{+}$ & $2^2S$ & 2.657 954 038 & 1.033 837 514 & 2.169 559 41 & 6.230 124 0 & 21.078 833 \\

 B$^{2+}$ & $2^2S$ & 3.408 499 326 & 0.760 963 580 & 1.132 751 15 & 2.301 051 6 & 5.509 837 4 \\
       \hline          
 atom/ion  & state & $\langle r^{-1}_{ee} \rangle$ & $\langle r_{ee} \rangle$ & $\langle r^2_{ee} \rangle$ &  $\langle r^3_{ee} \rangle$ &  $\langle r^4_{ee} \rangle$ \\
     \hline
 Li       & $2^2S$ & 0.732 736 059 & 2.889 506 202 & 12.283 005 19 & 64.032 401 8 & 385.287 08 \\

 Be$^{+}$ & $2^2S$ & 1.082 004 350 & 1.755 762 092 & 4.358 489 915 & 13.143 659 0 & 45.476 089 \\

 B$^{2+}$ & $2^2S$ & 1.426 153 105 & 1.278 851 329 & 2.275 177 620 & 4.895 786 45 & 12.056 833 \\
      \hline              
 atom/ion  & state & $\langle \frac12 p^2_{e} \rangle$ & $\langle \delta_{eN} \rangle$ & $\nu^{(a)}_{eN}$ & $\langle \delta_{ee} \rangle$ & $\nu^{(a)}_{ee}$ \\ 
     \hline
 Li        & $2^2S$ & 2.492 685 087 & 4.614 201 2 & -2.999 79 & 0.181 553 & 0.498 0 \\

 Be$^{+}$  & $2^2S$ & 4.774 894 384 & 11.701 015 & -3.997 59 & 0.528 005 & 0.485 2 \\

 B$^{2+}$  & $2^2S$ & 7.808 151 989 & 23.828 655 & -5.012 87 & 1.162 541 & 0.475 7 \\
     \hline\hline
  \end{tabular}
  \end{center}
The expected value of the electron-nucleus cusp $\overline{\nu}_{eN}$ for these atomic systems are -3.0, -4.0 and -5.0, respectively. The expected value of the electron-electron cusp equals 0.5 for all systems.
   \end{table}
%

% Table 2 

\begin{table}[tbp]
   \caption{The total energies $E$ (in atomic units) of some bound ${}^{2}S(L = 0)-$states of three-electron atomic systems 
            calculated with the use of six-dimensional gaussoids.}
     \begin{center}
%     \scalebox{0.72}{%
     \begin{tabular}{| c | c | c | c |}
      \hline\hline
   atom ( state ) & $E$     & ion (state) &  $E$ \\
        \hline
 Li($2^2S$-state) & -7.478 059 458 7 & Be$^{+}$ ($2^2S$-state) & -14.324 762 515 \\

 Li($3^2S$-state) & -7.354 097 071 4 & B$^{2+}$ ($2^2S$-state) & -23.424 605 665 \\ 

 Li($4^2S$-state) & -7.318 370 721 7 & C$^{3+}$ ($2^2S$-state) & -34.775 510 611 \\   

 Li($5^2S$-state) & -7.303 255 727 0 & N$^{4+}$ ($2^2S$-state) & -48.376 895 985 \\ 

 Li($6^2S$-state) & -7.294 895 144 5 & O$^{5+}$ ($2^2S$-state) & -64.228 536 815 \\  
      \hline\hline
  \end{tabular}
  \end{center}
  \end{table}
%

% Table 3

 \begin{table}[tbp]
   \caption{Calculation of properties using the six-dimensional gaussoids. 
            The expectation values of a number of electron-nuclear ($en$) and electron-electron ($ee$) 
            properties of some $n^3S-$states of the ${}^{\infty}$Li atom (in atomic units).}
     \begin{center}
%     \scalebox{0.72}{%
     \begin{tabular}{| c | c | c | c | c | c | c | c |}
       \hline\hline          
 atom/ion  & state  & $\langle r^{-2}_{eN} \rangle$ & $\langle r^{-1}_{eN} \rangle$ & $\langle r_{eN} \rangle$ & $\langle r^2_{eN} \rangle$  & $\langle r^3_{eN} \rangle$  &  $\langle r^4_{eN} \rangle$ \\
     \hline
       Li & $2^2S$ & 10.080 300 60 & 1.906 039 33 & 1.663 121 31 & 6.117 329 34 & 30.8544 & 183.163 \\

       Li & $3^2S$ &  9.981 911 41 & 1.837 281 71 & 3.762 215 05 & 39.369 603 & 494.5408 & 6765.5 \\

       Li & $4^2S$ &  9.963 898 91 & 1.816 606 44 & 6.774 635 06 & 138.026 11 & 3172.75 & 76948.0 \\

       Li & $5^2S$ &  9.958 334 21 & 1.807 613 45 & 10.78 694 9 & 363.470 48 & 13448.4 & 519573. \\

       Li & $6^2S$ &  9.959 661 84 & 1.804 788 25 & 14.688 387 & 704.528 1 & 37165.4 & $\approx$ 2054280.\\
               \hline 
 Be$^{+}$ & $2^2S$ & 18.998 594 7 & 2.657 962 911 & 1.033 794 655 & 2.169 285 0 & 6.228 678 0 & 21.071 451 \\

 B$^{2+}$ & $2^2S$ & 30.753 962 5 & 3.408 508 336 & 0.760 947 632 & 1.132 688 0 & 2.300 819 5 & 5.5088 155 \\

 C$^{3+}$ & $2^2S$ & 45.344 328 9 & 4.158 733 353 & 0.604 265 284 & 0.699 598 0 & 1.103 077 7 & 2.050 513 \\
       \hline          
 atom/ion  & state & $\langle r^{-2}_{ee} \rangle$ & $\langle r^{-1}_{ee} \rangle$ & $\langle r_{ee} \rangle$ & $\langle r^2_{ee} \rangle$ &  $\langle r^3_{ee} \rangle$ &  $\langle r^4_{ee} \rangle$ \\
     \hline
 Li &  $2^2S$ & 1.460 399 80 & 0.732 740 96 & 2.889 359 31 & 12.280 848 13 & 64.0027 & 384.859 \\

 Li &  $3^2S$ & 1.384 490 26 & 0.609 100 81 & 7.062 975 7 & 78.736 745 89 & 995.118 & 13650.0 \\
 
 Li &  $4^2S$ & 1.370 249 96 & 0.570 069 68 & 13.080 897 & 276.071 50 & 6356.9 & 154310. \\

 Li &  $5^2S$ & 1.366 046 05 & 0.552 577 36 & 21.102 651 & 726.956 5 & 26915. & 1040300. \\ 

 Li &  $6^2S$ & 1.365 389 62 & 0.546 446 79 & 28.904 765 & 1409.07 & 74356. & 4110650. \\
         \hline 
 Be$^{+}$ & $2^2S$ & 2.965 647 81 & 1.082 010 036 & 1.755 676 045 & 4.357 936 84 & 13.140 717 & 45.460 84 \\

 B$^{2+}$ & $2^2S$ & 5.003 262 87 & 1.426 139 236 & 1.278 829 713 & 2.275 086 36 & 4.895 417 0 & 12.054 932 \\

 C$^{3+}$ & $2^2S$ & 7.573 737 42 & 1.768 726 776 & 1.009 280 701 & 1.404 717 30 & 2.358 789 1 & 4.526 4953 \\
      \hline              
 atom/ion  & state & $\langle \frac12 p^2_{e} \rangle$ & $\langle \frac12 p^2_{N} \rangle$ & $\langle \delta_{eN} \rangle$ & $\langle \delta_{ee} \rangle$ &  &  \\ 
     \hline
   Li &  $2^2S$ & 2.492 690 62 & 7.779 909 06 & 4.606 397 & 0.181 679 &  &  \\

   Li &  $3^2S$ & 2.451 380 14 & 7.646 130 68 & 4.561 635 & 0.179 148 &  &  \\

   Li &  $4^2S$ & 2.440 297 94 & 7.610 871 32 & 4.539 129 & 0.178 976 &  & \\ 

   Li &  $5^2S$ & 2.435 882 57 & 7.594 940 22 & 4.509 019 & 0.179 340 &  &  \\ 
 
   Li &  $6^2S$ & 2.434 383 16 & 7.591 249 73 & 4.503 630 & 0.178 076 &  &  \\
         \hline
  Be$^{+}$ & $2^2S$ & 4.774 920 780 & 14.777 678 465 & 11.675 957 & 0.527 406 &  &  \\

  B$^{2+}$ & $2^2S$ & 7.808 200 841 & 24.030 676 084 & 23.764 692 & 1.159 936 &  &  \\

  C$^{3+}$ & $2^2S$ & 11.591 836 456 & 35.535 670 475 & 42.252 507 & 2.166 278 &  &  \\
     \hline\hline
  \end{tabular}
  \end{center}
  \end{table}
%

% Table 4

\begin{table}
\caption{Calculated Hy-CI / CI energies of the ground S-state and low-lying S-, P-, D-, F-, G-, H-, I-, K-, L-, M-, 
and N-excited states of the C$^{3+}$ and F$^{6+}$ ions.}
\begin{center}
%\scalebox{0.75}{
\begin{tabular}{| l | l | l | l | l |}
\hline\hline
 State \quad & \qquad Energy (Li)$^a$ \cite{Bel1} \qquad & \qquad Energy (Be$^{+}$)$^b$ \cite{Bel1} \qquad & \qquad Energy (C$^{3+}$)$^c$ \qquad & \qquad  Energy (F$^{6+}$)$^d$ \\
\hline\hline 
 2$^2$S & -7.478 058 969 & -14.324 761 678 & -34.775 407 123   &  -82.330 336 543   \\
 2$^2$P & -7.410 149 407 & -14.179 327 999 & -34.482 061 251   &  -81.820 872 700   \\
\hline
 3$^2$S & -7.354 093 706 & -13.922 784 968 & -33.396 193 013   &  -78.441 024 591   \\
 3$^2$P & -7.337 113 114 & -13.885 115 345 & -33.317 932 151   &  -78.300 897 021   \\
 3$^2$D & -7.335 512 623 & -13.878 041 021 & -33.296 030 365   &  -78.255 989 200   \\
\hline
 4$^2$S & -7.318 517 759 & -13.798 706 849 & -32.947 562 660   &  -77.139 946 952   \\
 4$^2$P & -7.311 811 529 & -13.783 574 124 & -32.915 327 522   &  -77.083 158 097   \\
 4$^2$D & -7.311 211 047 & -13.780 663 883 & -32.906 635 904   &  -77.063 937 284   \\
 4$^2$F & -7.310 610     & -13.779 946     & -32.905 543       &  -77.062 191       \\
\hline
 5$^2$S & -7.303 512 964 & -13.744 580 355 & -32.746 301 042   &  -76.542 595 063  \\
 5$^2$P & -7.300 137 068 & -13.736 854 458 & -32.730 766 119   &  -76.521 194 900  \\
 5$^2$D & -7.299 889 424 & -13.735 539 056 & -32.725 744 797   &  -76.512 144 229  \\
 5$^2$F & -7.299 460     & -13.735 055     & -32.725 670       &  -76.511 079      \\
 5$^2$G & -7.299 430     & -13.734 968     & -32.725 629       &  -76.511 030      \\
\hline
 6$^2$S & -7.295 739 603 & -13.716 223 859 & -32.640 301 515   &  -76.234 472 428  \\
 6$^2$P & -7.293 967 122 & -13.711 935 268 & -32.631 103 514   &  -76.217 166 129  \\
 6$^2$D & -7.293 697 654 & -13.710 5       & -32.627 8         &  -76.211 61       \\           
 6$^2$F & -7.293 323     & -13.710 6       & -32.627 9         &  -76.211 61       \\          
 6$^2$G & -7.293 32      & -13.710 6       & -32.627 6         &  -76.138 16       \\          
 6$^2$H & -7.293 32      & -13.710 6       & -32.627 9         &  -76.211 6        \\          
\hline
 7$^2$S & -7.291 231 582 & -13.699 298 491 & -32.576 281 553   &  -76.044 227 651  \\
 7$^2$P & -7.289 814 402 & -13.696 36      & -32.570 511 575   &  -76.033 783 122  \\
 7$^2$D & -7.289 8       & -13.695 4       & -32.568 5         &  -76.030 6        \\    
 7$^2$F & -7.289 6       & -13.695 8       & -32.568 9         &  -76.031 0        \\    
 7$^2$G & -7.289 6       & -13.695 7       & -32.568 7         &  -75.929 0        \\    
 7$^2$H & -7.289 6       & -13.695 8       & -32.568 9         &  -76.031 0        \\     
 7$^2$I & -7.289 6       & -13.695 8       & -32.568 9         &  -76.031 0        \\
\hline\hline
\end{tabular}
%}
\end{center}
\end{table}

\begin{table}
{\bf Continuation Table IV} Calculated Hy-CI / CI energies of the ground S-state and low-lying S-, P-, D-, F-, G-, H-, I-,
K-, L-, M-, and N-excited states of the C$^{3+}$ and F$^{6+}$ ions. \\
\begin{center}
%\scalebox{0.75}{
\begin{tabular}{| l | l | l | l | l |}
\hline\hline
 State \quad & \qquad Energy (Li)$^a$ \cite{Bel1} \qquad & \qquad Energy (Be$^{+}$)$^b$ \cite{Bel1} \qquad & \qquad Energy (C$^{3+}$)$^c$ \qquad & \qquad  Energy (F$^{6+}$)$^d$ \\
\hline\hline 
 8$^2$S & -7.288 393 829 & -13.688 174 464 & -32.534 198 075   &  -75.918 879 780  \\     
% 8$^2$P & -7.285 550     & -13.680 042 709 & -32.528 804 419   &  -75.835 137 409  \\                      
% 8$^2$D & -7.284 196     & -13.677 926     & -32.324 485       &  -75.831 120      \\                      
 8$^2$F & -7.287 0       & -13.685 9       & -32.530 4         &  -75.913 6        \\              
 8$^2$G & -7.287 2       & -13.686 2       & -32.530 6         &  -75.913 6        \\
 8$^2$H & -7.287 2       & -13.685 2       & -32.530 5         &  -75.913 7        \\
 8$^2$I & -7.287 2       & -13.686 2       & -32.530 5         &  -75.913 7        \\
 8$^2$K & -7.287 2       & -13.686 3       & -32.530 6         &  -75.913 9        \\
\hline
 9$^2$L & -7.285 6       & -13.679 7       & -32.504 4         &  -75.833 5        \\
\hline
10$^2$M & -7.284 4       & -13.675 0       & -32.485 6         &  -75.776 0        \\
\hline
11$^2$N & -7.283 6       & -13.671 6       & -32.471 7         &  -75.733 5        \\
\hline
L=20    & -7.280 6       & -13.659 6       & -32.423 8         &  -75.585 6        \\
\hline\hline
\end{tabular}
%}
\end{center}
\footnotetext[1]{The ionization limit for the Li atom is -7.27991 34126 69305 96491 810(15) a.u.}
\footnotetext[2]{The ionization limit for the Be$^{+}$ ion is -13.65556 62384 23586 70207 810(15) a.u.}
\footnotetext[3]{The ionization limit for the C$^{3+}$ ion is -32.40624 660189 853031 055785(45) a.u. }
\footnotetext[4]{The ionization limit for the F$^{6+}$ ion is -75.53171 236395 949115(3) a.u. }
\end{table}

\begin{table}
\caption{Comparison of the energies of the 2$^2S$ ground and first 2$^2P$ excited states of the C$^{3+}$
and F$^{6+}$ ions calculated by the Hy-CI in this work and by Hy method \cite{YTD}. Energies in a.u.}
\begin{center}
\begin{tabular}{| c | c | c | c | c |}
\hline\hline
  Ion     &  State &   This work       &  Ref. \cite{YTD} \\
\hline
C$^{3+}$  & 2$^2S$ & -34.775 407 123   &  -34.775 511 275 626(12) \\
C$^{3+}$  & 2$^2P$ & -34.482 061 251   &  -34.482 103 179 278(33) \\
\hline
F$^{6+}$  & 2$^2S$ & -82.330 336 543   &  -82.330 338 097 298(12) \\
F$^{6+}$  & 2$^2P$ & -81.820 872 700   &  -81.820 880 913 294(30) \\
\hline\hline
\end{tabular}
\end{center}
\end{table}

\end{document}